\documentclass[sigplan,screen]{acmart}
\usepackage{booktabs,siunitx}
\usepackage{listings}
\usepackage{float}
\usepackage{multirow}
\usepackage{bbding}
\usepackage{pifont}
\usepackage{adjustbox}
\usepackage{rotating}
\usepackage{tcolorbox}
\usepackage{amsmath, amssymb, mathtools}
\usepackage{algorithm}
\usepackage{algpseudocode}
\usepackage{braket}
\usepackage{multicol}
\usepackage[normalem]{ulem}

\usepackage[inline]{enumitem}

\AtBeginDocument{%
  }

\setcopyright{none}
\copyrightyear{}
\acmYear{}
\acmDOI{}
\acmConference{}{}{}

\acmISBN{}

\newcommand{\FormulaSize}{\scriptsize}
\newcommand{\TableFontSize}{\scriptsize}
\newcommand{\shrink}[1][1]{\vspace{-#1\dimexpr0.1cm\relax}}

\usepackage{subcaption}

\lstdefinelanguage{OpenQASM}{
  morekeywords={OPENQASM,include,qubit,bit,measure,while,if,else,false,true,h,x,rx,rz,cx},
  sensitive=true,
  morecomment=[l]{//},
  morestring=[b]"
}

\begin{document}

\title[]{Mapping Dynamic, Hierarchical Quantum Circuits 
}

\author{Marouane Benbetka, Merwan Bekkar}
\affiliation{%
  \institution{New York University}
  \city{Abu-Dhabi}
  \country{UAE}
}
\email{mb10324@nyu.edu,mb10325@nyu.edu}

\author{Bokyeong Yoon}
\affiliation{%
  \institution{The Ohio State University}
  \city{Columbus}
  \country{USA}
 }
\email{yoon.731@osu.edu}

\author{Riyadh Baghdadi}
\affiliation{%
  \institution{New York University} 
  \city{Abu-Dhabi}
  \country{UAE}
}
\email{Baghdadi@nyu.edu}

\author{Martin Kong}
\affiliation{%
  \institution{The Ohio State University}
  \city{Columbus}
  \country{USA}
 }
\email{moreno.244@osu.edu}

\renewcommand{\shortauthors}{Benbetka, Bekkar, Yoon, Baghdadi and Kong}

\begin{abstract}

Qubit mapping is a critical pass in quantum compilation.
Despite various advances, dynamic circuits,
those exhibiting data dependent control-flow,
often resulting from qubit measurements, are not yet supported
by the vast majority of available qubit mappers.
The crucial limitation to overcome is 
the dependence on 
flat, one-dimensional representations of circuits.
Further, qubit mappers currently lack 
compiler abstractions that capture the hierarchical
nature of circuits, hindering the qubit mapping process.

In this paper,
\footnote{Accepted for publication to the 2027
International Symposium on Code Generation and Optimization (CGO).}
we introduce a new qubit mapping method and
analyses
to tackle hierarchical dynamic circuits. Our novelty resides in four key aspects: modeling (statically) sub-circuits in disjoint control-flow paths, introducing a novel Qubit Reconciliation
pass to maintain consistency between sub-circuit
and control-flow boundaries, a loop-entry remapping pass,
and a refined cost function enhanced for SWAP count,
circuit depth, circuit latency and error. 
We demonstrate the efficiency of our approach on a wide
range of dynamic circuits on two monolithic
Quantum Processing Units
of 127 and 156 qubits, and on chiplet hexagon-based QPUs.
On monolithic QPUs, our qubit mapper
improves the SWAP count by up to 52\%,
depth by up to 18\%, latency by up to 18.6\%,
and error by up to 40\%. On chiplet architectures,
we achieve improvements of up to 36\% on SWAP count,
8.7\% on depth, 15\% on latency, and 15\% of error.


\end{abstract}



\newcommand{\mk}[1]{\textcolor{red}{MK: #1}}
\newcommand{\mknew}[1]{\textcolor{red}{#1}}
\newcommand{\mkleft}[1]{\textcolor{red}{$\leftarrow$ MK: #1}}

\newcommand{\mb}[1]{\textcolor{green}{MB: #1}}
\newcommand{\mbnew}[1]{\textcolor{green}{#1}}
\newcommand{\mbleft}[1]{\textcolor{green}{$\leftarrow$ MB: #1}}

\newcommand{\rb}[1]{\textcolor{red}{RB: #1}}

\newcommand{\bm}[1]{\textcolor{blue}{BM: #1}}


\newcommand{\new}[1]{\textcolor{black}{#1}}
\newcommand{\old}[1]{}

\begin{abstract}

\end{abstract}

\maketitle

\newcommand{\ourtool}[0]{{\tt DynamiQ}}

\makeatletter
\patchcmd{\ALG@beginalgorithmic}{\itemsep\z@}{\itemsep=0.3ex plus0.1ex}{}{}
\makeatother

\algrenewcommand\algorithmicindent{0.8em}

\algrenewcommand{\algorithmiccomment}[1]{\hfill{\scriptsize\textit{// #1}}}

\setlength{\textfloatsep}{5pt plus 1pt minus 2pt}
\setlength{\intextsep}{5pt plus 1pt minus 2pt}


\section{Introduction}
\label{sec:intro}

Recent advances in technology have enabled 
mid-circuit measurements and qubit reset in quantum
systems
\cite{exploiting-dynamic-qc.physrevlet.2021,dynamic-circuits-honeywell.nature.2021}.
Such features are fundamental in enabling new classes
of experimental methods such as Repeat-Until-Success, 
classical feed-forward control,
Quantum Error Correction, and Fault-Tolerant Quantum Computing.
This paper tackles the problem of qubit mapping and routing
\old{
for dynamic circuits (circuits with data-dependent control-flow), where mid-circuit measurements
are leveraged to condition the execution of gates forming sub-circuits.
}
\new{
for circuits with nested, data-dependent control-flow, where the result of mid-circuit measurements
are used to select the execution of gates forming sub-circuits.
}
We target classes of circuits that exhibit
selective  and iterative (loop-based) control-flow
constructs with
fixed-count, symbolic-count and nested control-flow structures.

Dynamic control-flow is the principal inhibitor of
current qubit mapping methods, as it induces execution
uncertainty and a dynamic dependence structure in quantum circuits.
Multiple (new) classes of applications use
mid-circuit measurements to condition the execution. 
Examples include
teleportation and fault-tolerant protocols, 
{\em magic state distillation}, and phase correction.
Teleportation protocols leverage
classical feedback 
using circuits of constant depth that generate multi-partite entanglement \cite{multi-part-entanglement-with-classical-feedback.arxiv.2025}.
Additionally, dynamic control-flow also
enables fault-tolerant protocols, which often rely
on feed-forward mechanisms. 

There is a critical demand for effective qubit mapping
methods that support
{\em quantum circuits with data-dependent control-flow}
(dynamic circuits, for short),
as evidenced
by recent work \cite{adaptive-qc-advantage-exper.arvix.2023,cost-depth-prep-matrix-products-adaptive-quantum-circuits.aps.2024,singh-midcirc-correction.science.2023,magic-state-distil-rusp.arxiv.2023,multi-part-entanglement-with-classical-feedback.arxiv.2025,efficient-preparation-anyons-adaptive-qc.aps.2025,dynamic-adaptive-qaoa.physreva.2024,adaptive-circuits-with-feedback.aps.2023,adaptive-efficient-quantum-chemistry.rsc.2024,continuous-measurements.advphysx-2020,shallow-q-circuits-global-opt.2023,qubit-adapt-vqe.2021,exploiting-dynamic-qc.physrevlet.2021,adaptive-stabilization-qc,universal-qc-for-q-chem.2022,spectator-qubits.nature.2020,spectator-qubit-calibration.aps.2020,tracking-drift.nature.2020}.
While there has been a significant body of work
that maps and routes qubits to address nearest-neighbor
topological constraints 
\cite{siraichi.cgo.2018,dependence-aware-compilation-qec-annealing.oopsla.2025,wille.aspdac.2016,wille.dac.2019,mqt-qmap-jku-wille-burgholzer.ispd.2023,wille.aspdac.2014,sabre,olsq,cirq}, these have yet to tackle the problem in the context of dynamic
control-flow. The primary underlying challenge is that
the qubit mapping process is an inherently 
straight-line code compilation pass. To repair the connectivity
between qubit arguments of any two-qubit gate, 
the current logical-to-physical mapping must be known.
In light of this, naive mapping strategies for
loops with fixed -- known {\em a priori} -- iteration counts
resort to fully unrolling the loop body, effectively producing
a static circuit. However, {\tt for-loops} with parametric (symbolic) counts, {\tt while-loops} and {\tt branch}
conditionals with data-dependent control-flow (using classical bits obtained from mid-circuit measurements) pose
significant challenges to qubit mappers.


In this paper,
we introduce \ourtool{}, a novel multi-objective
qubit mapping and routing framework 
designed and implemented to
handle dynamic quantum circuits featuring data-dependent control-flow.
Unlike prior methods designed solely for static, acyclic circuits, \ourtool{} explicitly models control-flow structure, ensuring consistent mapping states across all possible execution paths.
It also introduces a fundamentally improved hardware-, error-, and latency-aware cost function.
\ourtool{}’s cost model accounts for the current circuit depth and integrates QPU calibration data (two-qubit error rates).
These additions make the mapping method
both error-resilient and depth-efficient, improving mapping quality for dynamic circuits.


The contributions of our work are:
i) A new hierarchical qubit mapping method
\new{
that unifies the qubit mapping state across
disjoint execution paths of the same nesting level by applying the token-swapping
approximation of~\cite{miltzow2016approximation} at control-flow boundaries}.
The new method supports 
fixed and symbolic iterative constructs and
multi-branch conditionals that exhibit data-dependent  control-flow;
ii) An enhanced multi-objective mapping cost function 
for straight-line circuits that builds on the Qlosure method~\cite{qlosure}. 
The new cost function
unifies four objectives, circuit depth, SWAP count, circuit error (i.e. aggregated error) and circuit latency;
iii) Due to the lack of quantum benchmark suites with dynamic control-flow, we develop a new generator for hierarchical
circuits, {\tt d-QUEKO}
(Dynamic-QUEKO) based on the QUEKO algorithm \cite{queko};
iv) Finally, an evaluation showing 
the effectiveness of \ourtool{} over the {\tt LightSABRE} mapper \cite{zou2024lightsabre}, 
on two monolithic QPUs (127/156-qubit) and two chiplet architectures, on
dynamic and rotated surface-code circuits.


 \shrink[2]
\section{Motivation}
\label{sec:motiv}
Numerous mapping methods have been proposed
to tackle the qubit mapping and routing
problem for {\em flat static circuits} \cite{sabre,siraichi.cgo.2018,wille.dac.2019,muqut,cirq,tket,qubit-mapping-hardware-aware.tqe.2020,fast-scalable-qubit-mapping.dac.2022,search-space-qc-qmap-jku.aspdac.2022,qmap-optimal-subarch-qmap.tqc.2023,jkq.iccad.2020,murali.asplos.2019,olsq,qubit-crosstalk.pra.2019,zulehner.date.2018}.
However, 
qubit mapping and routing for dynamic circuits 
(i.e. exhibiting
data-dependent control-flow conditions such as {\tt (b[0]==0})
remains
an open problem.
Work 
has been limited to 
producing a flattened program representation via loop
unrolling 
\cite{adaptive-qc-advantage-exper.arvix.2023}
and analyzing the 
Directed Acyclic Graph (DAG) 
representation
to achieve SWAP reduction by qubit reuse 
\cite{Hua.asplos.2023}.
Despite such advancements, mapping methods based solely on 
DAG analysis
lack the 
ability to handle hierarchical circuits.
This is because they lack composable 
compiler abstractions
that allow us to overcome the limitations inherent
in mapping straight-line circuits (those expressible as DAGs).
Compositional abstractions, such as those found
in affine frameworks \cite{qlosure,qrane} are key
to model, extract, and leverage circuit information outside
of a single circuit level.

\noindent
{\bf Data Dependent Control Conditions.}
A common assumption throughout 
qubit mapping methods is that all gates always execute. 
The presence of data-dependent conditions  that control subsets of gates
within iterative and selection constructs poses crucial challenges 
to mapping methods; namely, it introduces qubit mapping state gaps.
Not knowing the exact logical-to-physical qubit
mapping at any given point brings the mapping algorithm to a halt.
Such qubit mapping state gaps originate from 
disjoint control-flow paths, or different sequences of gates
being executed in each sub-circuit, leading to multiple qubit mapping states.

\noindent
{\bf Sub-Optimal Mappings Even without Multiple Control-Flow Paths.}
Qubit mapping methods, even those that target circuits represented
as DAGs, are currently limited to heuristics based on graph traversals,
which result in mapping decisions made with local information.
As a result, today's heuristics are still far from the optimal
solution \cite{olsq}.
Having multiple control-flow paths resulting from data-dependent conditions
further reduces the effectiveness of all qubit mapping methods, which assume
that all gates execute.

\noindent
{\bf The Key Contribution.}
Our key scientific contribution is a mapping algorithm
designed for hierarchical circuits that exploits
the transitive closure of dependence relations at each
level of the hierarchical circuit. This information
is used to prioritize mapping decisions. In addition, to
support nested and disjoint/distinct control-flow paths
taken by dynamic conditions
we leverage a novel {\em Qubit Reconciliation Pass}
whose task is to convert the qubit mapping state
between entry and exit points of hierarchical circuits
into a {\em known qubit steady state} at minimal cost.

\newcommand{\xmark}{\ding{55}}

\begin{table}[thb]
    \TableFontSize
    \caption{Support for mapping circuits with dynamic control-flow and multiple control-flow paths.
    Objective types: A*-Search (A*S), Bounded Distance (BD), Critical-Path Aware (CPA), Depth (D), Depth-Rate (DR), Error (E), Qubit Fidelity (QF), Swap Minimization (SM), Transitive Dependence Volume (TDV).
    Dyn. = Dynamic, Hier. = Hierarchical, Par. = Parametric, Ent. = Entry Remapping, Crit. = Critical Path.
    }
    \shrink[2]
    \label{tab:motiv}
    \centering
    \setlength{\tabcolsep}{2pt}
    \begin{tabular}{l|c|c|c|c|c|p{2cm}}
    \toprule
    {\bf Method} &
    {\bf Dyn.} &
    {\bf Hier.} &
    {\bf Par.} &
    {\bf Ent.} &
    {\bf Crit.} &
    {\bf Objectives}
    \\
    \midrule
    QISKIT LightSABRE \cite{sabre} &
    \checkmark &
    \checkmark &
    \xmark &
    \xmark &
    \xmark &
    E, D
    \\
    TKET \cite{tket} &
    \xmark &
    \xmark &
    \xmark &
    \xmark &
    \xmark &
    E, BD
    \\
    QMAP \cite{mqt-qmap-jku-wille-burgholzer.ispd.2023} &
    \xmark &
    \xmark &
    \xmark &
    \xmark &
    \xmark &
    BD, A*S, SM
    \\
    Cirq \cite{cirq} &
    \xmark &
    \xmark &
    \xmark &
    \xmark &
    \xmark &
    QF
    \\
    Qlosure \cite{qlosure} &
    \xmark &
    \xmark &
    \xmark &
    \xmark &
    \checkmark &
    TDV
    \\
    {\bf This work} &
    \checkmark &
    \checkmark &
    \checkmark &
    \checkmark &
    \checkmark &
    TDV, E, {\bf CPA, DR}
    \\
    \bottomrule
    \end{tabular}
\end{table}

We summarize the support of mapping methods in 
Tab.~\ref{tab:motiv}. Only QISKIT LightSABRE
\new{\cite{zou2024lightsabre}, the enhanced
SABRE method \cite{sabre} used by QISKIT,}
supports circuits that are both dynamic and hierarchical
(with nested control structures).
\new{
Mapping is performed recursively,
but optimizations cannot cross control-flow boundaries. \ourtool{} is different
from {\tt QISKIT's LightSABRE}
in four ways. First,
LightSABRE simply pads additional
branches with SWAPs to maintain the inherited
logical-to-physical qubit mapping. In
contrast, \ourtool{} maps each sub-circuit
and selects the one with the lowest mapping
cost as the steady-state, padding all other
branches with the needed SWAP operations.
Second, due to the dynamic nature,
LightSABRE adopts a {\em look through} 
look-ahead policy
that considers control-flow operations
as 1-qubit gates;
In this regard, \ourtool{} views
the remaining circuit as a static multi-qubit gate.
Third, our method considers the refinement
of the inherited mapping, allowing
to further improve the overall cost.
Finally, whereas LightSABRE prioritizes gates by how close their operands
currently are on the device, \ourtool{} 
weighs gates by how many
two-qubit operations depend on it, directly or indirectly, computed as a transitive closure over
the dependence relations at each level of the hierarchy.
}
Other mappers
do not support these features despite relying on several
types of objectives. None of the mappers support
qubit state remapping at
juncture points to better optimize across hierarchical
boundaries.
As we will see in the main evaluation, our 
updated cost function (Sec.~\ref{subsec:cost_function}) 
is quite effective in monolithic and modular QPUs,
and is capable of trading SWAP counts for error
in the presence of expensive QPU links.
\new{We also note that 
we enhance the 
{\tt Qlosure} mapping method \cite{qlosure} 
and use it to map static sub-circuits. 
The key differences w.r.t \cite{qlosure}
are the {\em qubit fidelity} and
{\em critical path} awareness extensions
described in Sec.~\ref{subsec:cost_function}
}

\begin{figure}[htbp]
    \centering

    \begin{subfigure}[b]{0.23\textwidth}
        \centering
        \includegraphics[width=\linewidth]{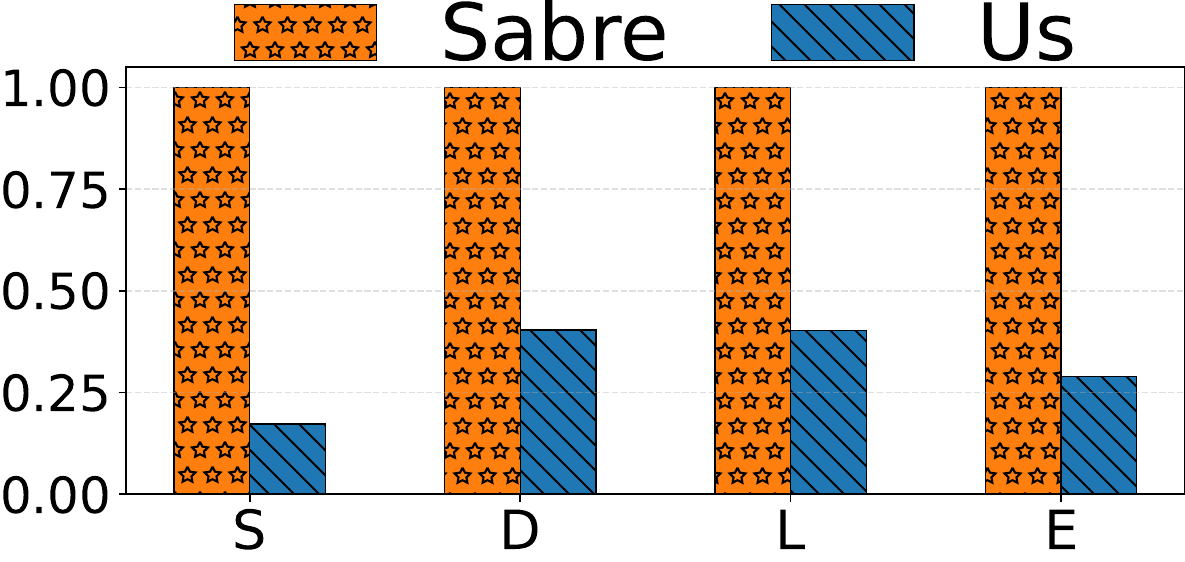} 
        \vspace{-4ex}
        \caption{
        81QBT\_5w\_5i} 
        \label{fig:motiv:dqueko:nested5}
    \end{subfigure}
    \hfill
    \begin{subfigure}[b]{0.23\textwidth}
        \centering
        \includegraphics[width=\linewidth,height=2cm]{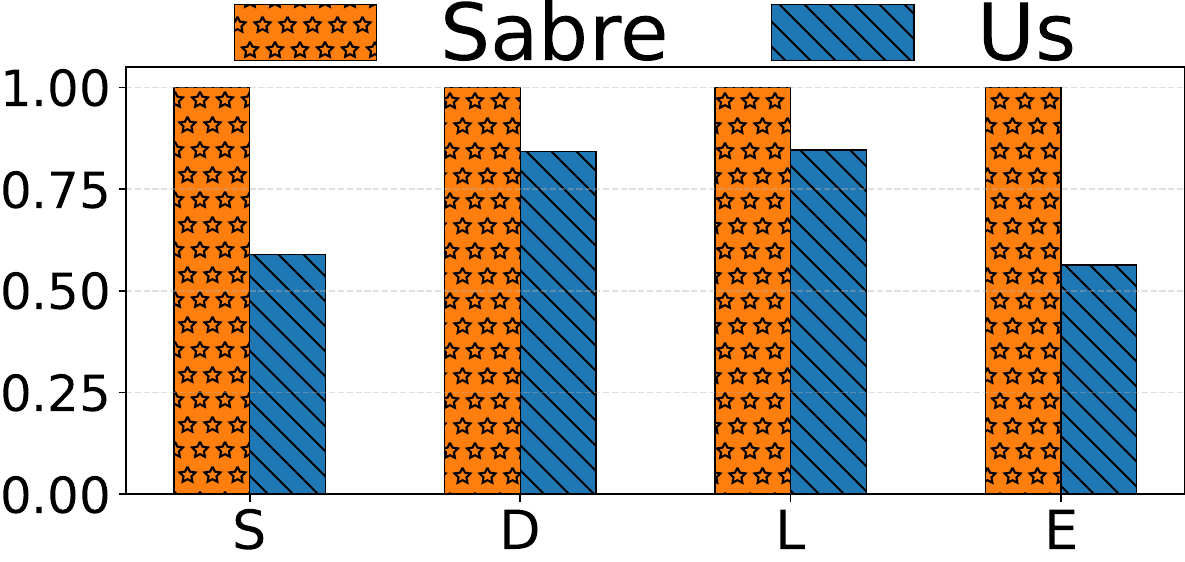}
           \vspace{-4ex}
        \caption{
        \footnotesize 121QBT\_60leaf\_depth} 
        \label{fig:motiv:dqueko:121qbt}
    \end{subfigure}

    \shrink[2]
    \caption{
    QISKIT Sabre vs. DynamiQ on d-QUEKO circuits (Sec.~\ref{sec:queko-loops}) on
    the IBM Brisbane QPU. Y-axis shows \emph{normalized values} 
    (lower is better) of four metrics: SWAPs (S), Depth (D), Latency (L), and Error (E). 
    Fig.~\ref{fig:motiv:dqueko:nested5}
    presents 5-nested loops, while
    the one on the right 
    Fig.~\ref{fig:motiv:dqueko:121qbt}
    presents five back-to-back loops
    interleaved with regular gates. 
    }
    \Description{
    Comparison between QISKIT LightSABRE and our method on the d-QUEKO benchmark on the IBM Brisbane backend. Figures present \emph{normalized values} for four metrics: SWAPs (S), Depth (D), Latency (L), and Error (E).
    }
    \label{fig:motivation-fig}
\end{figure}

\noindent
\begin{figure*}[ht]
    \centering
    \captionsetup[subfigure]{justification=centering, font=small, labelformat=parens, labelsep=space}

    \begin{subfigure}[t]{0.19\linewidth}
        \centering
        \includegraphics[width=1\linewidth]{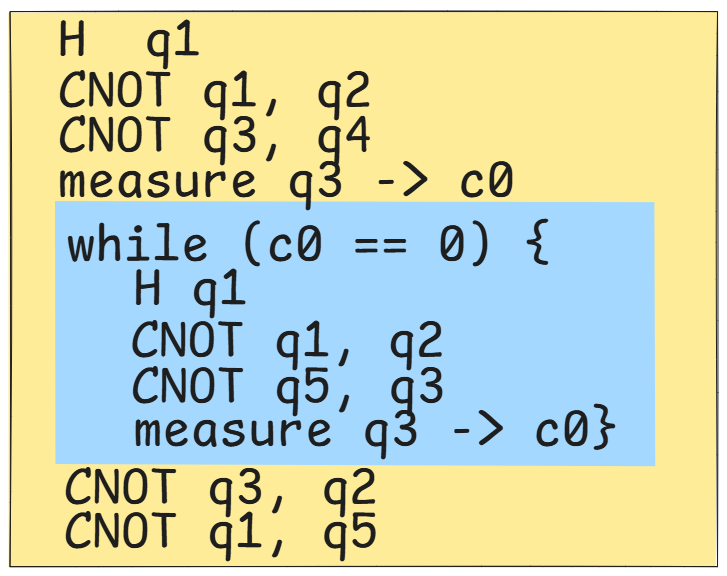}
        \begin{footnotesize}
        \vspace{-14px}
        \caption{Dynamic circuit.}
        \end{footnotesize}
        \label{fig:overview:input}
    \end{subfigure}
    \hspace{3em}
    \begin{subfigure}[t]{0.48\linewidth}
        \centering
        \includegraphics[width=1\linewidth]{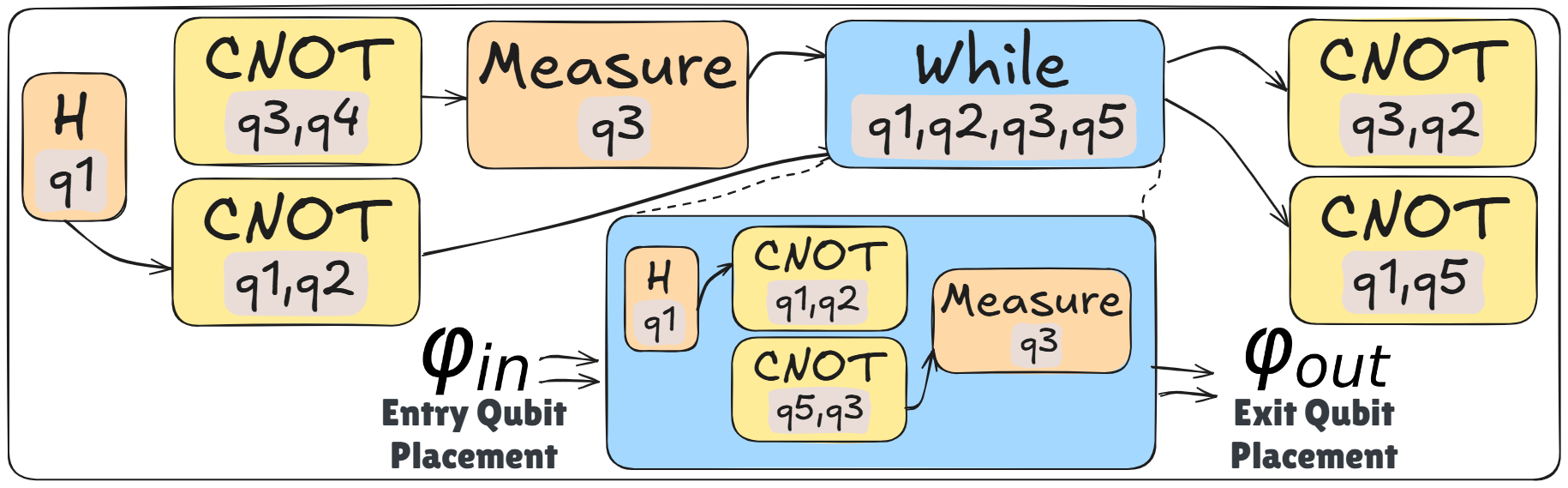}
        \begin{footnotesize}
        \vspace{-14px}
        \caption{Corresponding dependence graph.}
        \end{footnotesize}
        \label{fig:overview:qroqi-graph}
    \end{subfigure}
   \begin{footnotesize}
    \vspace{-10px}
    \caption{
  Dynamic circuits as hierarchical dependence graphs: 
dynamic blocks (loops, conditionals) form high-level nodes with subgraphs, 
compiled independently by \ourtool{} with fixed entry/exit mappings.
    }
    \label{fig:qroqi_dynamic_blocks}
   \end{footnotesize} 
   \vspace{-2ex}
   \Description{}
\end{figure*}

To conclude our motivation, Fig.~\ref{fig:motivation-fig} shows excerpts of our results.
Due to the lack of mappers capable of supporting
dynamic circuits, we compare our mapping method
with the LightSABRE implementation in QISKIT.
The effectiveness of our mapper is established
across four metrics: 
{\tt SWAP count} (S), circuit {\tt Depth} (D),
circuit {\tt Latency} (L),
and circuit {\tt Error} (E).
As we can see, our proposed method
achieves notable improvements on dynamic circuits, where
QISKIT LightSABRE is the only mapper capable of handling 
such circuits.

\section{Background}
\label{sec:bg}


\paragraph{Quantum Computing Fundamentals}

Quantum Processing Units (QPUs) 
compute using \emph{qubits}, 
which exist in a linear superposition of the computational basis states $|0\rangle$ and $|1\rangle$.
Qubit states are transformed via unitary operations known as \emph{quantum gates}.
Single-qubit gates (e.g., {\tt H}) act on individual qubits, while two-qubit gates (e.g., {\tt CNOT}, {\tt SWAP}) act on a qubit pair.
QPUs are composed of physical
qubits connected through a \emph{coupling graph} 
\(G_{\mathrm{hw}} = (Q_{\mathrm{phys}}, R_{\mathrm{hw}})\),
where each
vertex \(q_i \in Q_{\mathrm{phys}}\) is a qubit and each edge \((q_i, q_j) \in R_{\mathrm{hw}}\) denotes a
physical link. Two-qubit gates can be executed directly only if their operands share an edge in \(G_{\mathrm{hw}}\). 
\new{
Besides unitary gates, a \emph{measurement} operation
{\tt m(q,c)} 
collapses the state of a qubit $q$ onto the computational
basis and stores the outcome into a classical register $c$, where it stays readable
by the control electronics for the rest of the execution~\cite{cross2022openqasm}. 
}

{\bf Gate Errors and Fidelity.}  
In current QPUs,
each gate has an \emph{error rate} \(\varepsilon\),
often expressed through its \emph{fidelity} \(F = 1 - \varepsilon\), where \(\varepsilon\) denotes the probability of failure.  
As each link \((q_i,q_j)\) in the graph $G_\mathrm{hw}$ has its own fidelity, 
differentiating among them leads to significant qualitative
improvements.
Moreover, two-qubit gates are an order of magnitude noisier than single-qubit operations, and dominate the overall circuit fidelity, making topology- and fidelity-aware compilation crucial for reliable execution.

{\bf Dynamic Circuits and Classical Control.}  
Recent QPUs support \emph{dynamic circuits}, which allow 
to condition gates
to 
measurement results.  
This enables data-dependent control through constructs such as \texttt{if–else} branches, \texttt{while} loops, and conditional gate execution.  
For instance, a qubit 
measurement 
can
control
the execution of a later gate 
or 
the 
execution of a repeat-until loop
once a measurement outcome is observed.
\new{
Such feedback mechanisms make 
permit
to implement adaptive algorithms, error mitigation routines, and mid-circuit corrections directly on hardware.
Fig.~\ref{fig:qroqi_dynamic_blocks}a illustrates the latter case: qubit $q_3$ is
measured into the classical bit $c_0$, and the loop body repeats while $c_0 = 0$.
Since the body re-measures $q_3$, the number of iterations is decided at run-time
and is unavailable to the compiler.
}

{\bf Quantum Routing Problem.}\label{subsec:routing-problem-definition}  
Quantum circuits are defined over \emph{logical qubits}, while hardware executes operations on \emph{physical qubits}.  
The compiler maintains a mapping \(\varphi: Q_{\mathrm{logical}} \rightarrow Q_{\mathrm{phys}}\) that assigns each logical qubit to a physical one on the device.  
A two-qubit gate acting on logical qubits \((q_i, q_j)\) is \emph{feasible} only if the corresponding physical qubits \((\varphi(q_i), \varphi(q_j))\) share an edge in \(G_{\mathrm{hw}}\).  
If not, the compiler inserts \emph{SWAP gates} along a path to move qubit states until the pair becomes adjacent.

\noindent\textbf{Example:}  
Consider a linear hardware topology with physical qubits \(p_0, p_1, p_2, p_3\) connected as  
\(p_0{-}p_1{-}p_2{-}p_3\).  
Suppose a CNOT acting on qubits \(q_A\) and \(q_D\), initially mapped to \(p_0\) and \(p_3\), which are not directly connected.  
To make them adjacent, the compiler applies a sequence of SWAPs: SWAP\((p_0, p_1)\), and
SWAP\((p_1, p_2)\). 
This makes the logical qubits \(q_A\) (in p2) and \(q_D\) adjacent, allowing the CNOT to be executed.

The \emph{quantum routing problem} consists
of determining where and when to insert 
SWAP gates 
that make operands of two-qubit gates physically adjacent,
while minimizing the circuit depth and noise, avoiding
swaps on low-fidelity edges or moves that place qubits on noisy links, as such choices increase circuit errors.

 \shrink[2]
\section{Overview}
\label{sec:overview}

We now describe how \ourtool{} handles dynamic circuits.
We lift every quantum program into 
\old{\em an affine } 
\new{\em a hierarchical}
dependence graph,
where each gate, measurement, or control-flow construct is modeled as a node. Edges capture 
dependencies between gates sharing at least one logical qubit.
Fig.~\ref{fig:qroqi_dynamic_blocks} 
shows this process
in a circuit with a 
\texttt{while}-loop; the loop body is abstracted as a high-level node in the outer DAG,
but internally it represents a subgraph corresponding to one iteration of the loop.
In this hierarchical view, a dynamic block behaves as 
a {\em multi-qubit gate}
(of all accessed qubits inside the block).
For example, a loop body containing gates 
{\tt CNOT($q_1,q_2$)} and {\tt CNOT($q_5,q_3$)} 
is represented as a single node that uses
qubits $\{q_1, q_2, q_3, q_5\}$, thereby preserving dependence edges between the block
and other operations.

\begin{figure*}[t]
    \shrink[1]
    \centering
    \captionsetup[subfigure]{justification=centering, font=small, labelformat=parens, labelsep=space}

    \begin{subfigure}[b]{0.45\linewidth}
        \centering
        \includegraphics[width=1\linewidth]{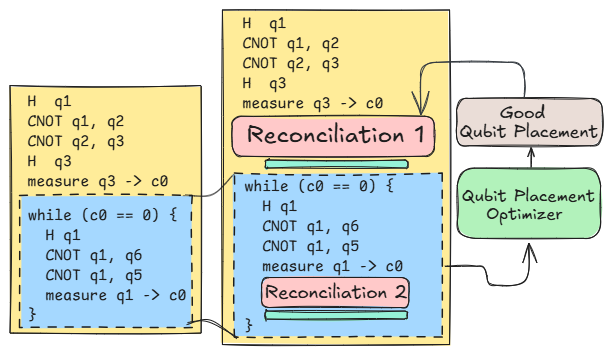}
        \vspace{-4ex}
        \caption{Loop body extraction and reconciliation.}
        \label{fig:qroqi_loop_mapping}
    \end{subfigure}
    \begin{subfigure}[b]{0.42\linewidth}
        \centering
        \includegraphics[width=1\linewidth]{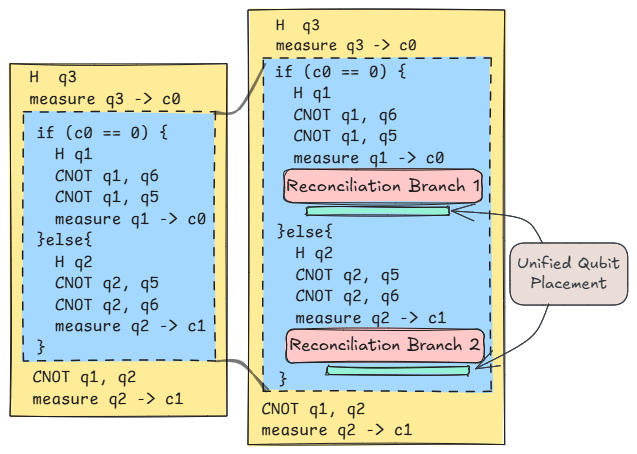}
        \vspace{-4ex}
        \caption{Conditional branches and merge reconciliation.}
        \label{fig:qroqi_conditional_mapping}
    \end{subfigure}
    \vspace{-10px}
    \caption{
Reconciliation strategies in \ourtool{}: 
(a) loops learn an entry mapping $\varphi_{\mathrm{in}}^{\star}$ for consistent placement; 
(b) conditionals reconcile branch mappings at merge points for deterministic execution.
    }
    \label{fig:qroqi_loop_conditional_combined}
    \Description{}
\vspace{-2ex}
\end{figure*}

{\bf Mapping Strategy for Dynamic Blocks.}
Because 
execution paths in dynamic circuits cannot
be determined at compile-time,
\ourtool{} imposes a key constraint:
\textbf{each dynamic block is compiled with a fixed entry and exit qubit placement.}
Every block $B$ is associated with a deterministic pair
$(\varphi_{\mathrm{in}}^B,\, \varphi_{\mathrm{out}}^B)$,
where $\varphi_{\mathrm{in}}^B$ denotes the \emph{logical-to-physical qubit mapping}
at the point of entry into the block (i.e., how logical qubits are placed on hardware before the block executes),
and $\varphi_{\mathrm{out}}^B$ denotes the corresponding mapping
upon exiting the block after compilation.
This invariant ensures that, regardless of how many loop iterations execute
or which branch is taken, 
the compiler always returns to a 
predictable placement
state, enabling deterministic continuation of the mapping process.

To enforce these invariants, \ourtool{} inserts 
{\em reconciliation passes} that restore a target mapping between blocks:
for a loop, a reconciliation pass 
is invoked for the sub-circuit associated to each distinct control-flow path
at the end of each iteration
so that execution always returns to $\varphi_{\mathrm{in}}$; 
for a {conditional block}, reconciliation gates are appended to each
branch to merge all exit mappings into a common $\varphi_{\mathrm{out}}$.
More details are given in Sec.~\ref{subsec:routing_algorithm}.

The \ourtool{} mapping algorithm  
operates recursively:
whenever it encounters a dynamic node in the DAG, 
\ourtool{} recurses into the node’s
sub-graph (e.g., the while-loop in Fig.~\ref{fig:qroqi_dynamic_blocks}), 
applies its mapping and reconciliation logic, 
and then resumes 
global routing with a known consistent placement.
At runtime, whichever branch or iteration path the quantum circuit follows,
the precomputed SWAPs ensure that
qubit placements remain valid.

\shrink[2]
\section{Mapping Dynamic Circuits}
\label{sec:mapping}

\newcommand{\phistar}[0]{\varphi^{\star}_{in}}
\newcommand{\typedvarphi}[1]{\varphi_{\mathrm{#1}}}
\newcommand{\smap}[1]{\mathcal{S}_{\text{#1}}}


\subsection{Considerations of Mapping Control-Flow Blocks}
\label{sec:mapping:considerations}

\begin{figure}[ht]
    \shrink[2]
    \centering
    \includegraphics[width=0.99\columnwidth]{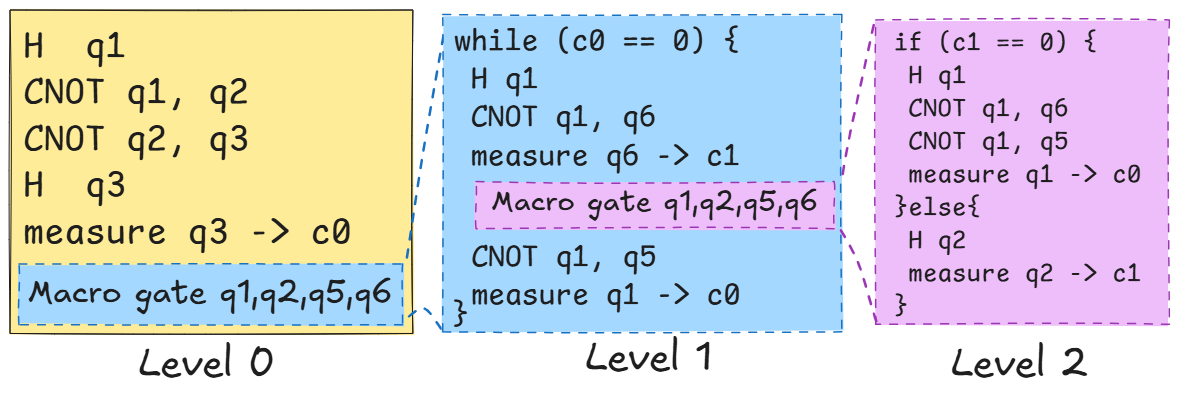}
    \shrink[5]
    \caption{
Handling of nested dynamic constructs in \ourtool{}.
Inner regions are compiled as independent blocks and
abstracted as \emph{macro gates} for outer routing.
    }
    \Description{
    Handling of nested dynamic constructs in \ourtool{}.
Inner regions are compiled as independent blocks and
abstracted as \emph{macro gates} for outer routing, enabling recursive compilation.
    }
    \label{fig:qroqi_nested_blocks}
    \shrink[2]
\end{figure}

\ourtool{} 
supports structured quantum programs 
using loops and conditional statements. 
Our approach 
works entirely at compile-time, and
supports nested control-flow structures, including fixed-count and parametric loops ({\tt for-loops} and {\tt while-loops}).
\new{
Hardware-wise, \ourtool{} targets QPUs whose qubit connectivity is 
defined by a
coupling graph $G_{\mathrm{hw}}(Q_{\mathrm{phys}}$, $R_{\mathrm{hw}})$. Per-edge two-qubit error rates and per-gate durations are optional refinements that default to constants when calibration data is unavailable.
}
\new{Further, \ourtool{} preserves the inherited
control structure: program gates never move
across region boundaries and regions are never fused. Each region's qubit footprint is
therefore fixed by the source program. Transformations that restructure control-flow, such
as fusing adjacent regions, are complementary and could be applied before \ourtool{} runs.
We also do not unroll loops,
which can increase the precision
of our analysis
when the trip-count is known at compile-time.
}
Fig.~\ref{fig:qroqi_nested_blocks} shows
an exemplar quantum circuit handled by our
framework.
\new{
Tab.~\ref{tab:glossary} summarizes the notation used throughout this section.
}

\label{subsubsec:loop_constructs}


{\bf Iterative Constructs (Loops).}
When \ourtool{} encounters a loop, the loop body is extracted as an independent
sub-circuit of the program and compiled separately (example in
Fig.~\ref{fig:qroqi_loop_mapping}). \ourtool{}
proceeds
\emph{recursively} 
on each new conditional or iterative construct.

Mapping loops with quantum operations demands
knowing the mapping state $\typedvarphi{in}$ at the
loop entry
in order to produce the sequence of SWAP operators
$\mathcal{S}_{body}$ and the final state $\typedvarphi{out}$
resulting from the mapping (updated from $\typedvarphi{in}$). However, to preserve
the 
consistency of qubit mappings across loop iterations
it is necessary to return to a known qubit state
before proceeding to the next iteration. 
The traditional approach to this in the QISKIT
compiler is to use $\mathcal{S}^{-1}_\textrm{body}$
(the same SWAPs applied in reverse order)
to return
to the known state $\typedvarphi{in}$.

\begin{table}[h!]
\centering

\setlength{\tabcolsep}{3pt}
\renewcommand{\arraystretch}{0.9}
\TableFontSize

\caption{Definitions of symbols used in later sections.
}
\label{tab:glossary}
\vspace{-3ex}

\begin{tabular}{@{}c p{6.5cm}@{}}
\toprule
\textbf{Notation} & \textbf{Definition} \\
\midrule

$q_i, p_j$ & Logical qubit $i$ and physical qubit $j$. \\

$Q_{\mathrm{logical}}, Q_{\mathrm{phys}}$ & Sets of logical and physical qubits. \\

$\varphi$ & Mapping from $Q_{\mathrm{logical}}$ to $Q_{\mathrm{phys}}$. \\

$G_{\mathrm{hw}}(Q_{\mathrm{phys}}, R_{\mathrm{hw}})$ & Hardware coupling graph. \\

$R_{\mathrm{hw}}$ & Directed edges of $G_{\mathrm{hw}}$. \\

$D_{\mathrm{phys}}[p_i,p_j]$ & Shortest-path distance between pi and pj. \\

$B$ & Dynamic block (loop or conditional branch). \\


$\omega_g$ & Dependence weight of gate $g$. \\

$S$ & Sequence of SWAP gates. \\

$S^{-1}$ & SWAP sequence obtained by reversing $S$, used to undo its effect. \\

$C$ & Emitted gate sequence. \\

\bottomrule
\end{tabular}
\end{table}

However, this strategy
can be highly inefficient if the inherited mapping is of poor quality. 
To address this, \ourtool{} 
applies a \emph{bidirectional optimization process} to 
find a
high-quality entry placement before fixing it 
(See Sec.~\ref{subsec:routing_algorithm}).

{\bf Conditional Constructs.}
Branches of conditionals are extracted as 
independent sub-circuits fed to \ourtool{},
recursively.
(See
Fig.~\ref{fig:qroqi_conditional_mapping}).
Branches are optimized separately to
generate valid hardware-aware mappings.
The mapping obtained at
the end of each branch represents the final placement 
$\typedvarphi{out}^{B_i}$
that would result if that
specific control path were executed.
Then, since all branches must eventually merge, 
\ourtool{} evaluates all resulting
branch mappings using the same cost model described in Sec.~\ref{subsec:cost_function}.
Among these branches, the mapping with
lowest overall cost is chosen as the \emph{unified post-branch mapping},
denoted $\varphi_{\mathrm{merge}}$.
\ourtool{}
generates reconciliation passes for all branches to ensure convergence at the
merge point. For each branch $B_i$, a minimal sequence of SWAP operations is
computed—using the token-swapping method of
Sec.~\ref{subsec:reconciliation}—to transform the branch’s final
mapping $\varphi_{\mathrm{out}}^{B_i}$ into $\varphi_{\mathrm{merge}}$. Each
reconciliation sequence is appended to the end of its corresponding branch,
guaranteeing that all branches terminate in the same qubit configuration.

As shown
in Fig.~\ref{fig:qroqi_conditional_mapping}, 
this process
introduces a reconciliation stage for all branches
(\emph{Branch 1} 
and \emph{Branch 2}), to converge
to a 
unified
placement, later propagated forward for global consistency.
This ensures deterministic routing and depth-efficient
execution without requiring runtime evaluation of conditional paths.


{\bf Nested Dynamic Constructs.}
Nested dynamic control-flow regions
are compiled
independently with clearly defined entry and exit mappings,
$\typedvarphi{in}$ and $\typedvarphi{out}$.
Because sub-circuits
are reconciled to a known final placement, their
exit configuration can be directly
reused as the entry mapping of any enclosing or subsequent region. This design
enables seamless integration of arbitrarily nested dynamic structures, such as
loops within loops or conditionals inside iterative regions.

Fig.~\ref{fig:qroqi_nested_blocks} shows an example nesting pattern in which dynamic blocks
appear hierarchically, such as an \texttt{if-else} structure nested inside a
\texttt{while} loop. At the outer level, the loop body (Level~1) includes an
inner control-flow region (Level~2) that itself behaves as a dynamic block. Each
level maintains its own entry and exit mappings, reconciled locally through the
mechanism described in 
Sec.~\ref{subsec:reconciliation}.

The compiler recursively
invokes \ourtool{} on a sub-block 
to
determine the exit placement $\varphi_{\mathrm{out}}^B$. 
Compilation proceeds as if $B$ were a single static gate that transforms $\typedvarphi{in}^{B}$ into $\typedvarphi{out}^{B}$.



\subsection{IR Extraction}
\label{subsec:ir:short}
\ourtool{} represents circuits with a hierarchical IR.
At any level, individual gates with fixed qubit accesses and macro-gates (e.g., loops or conditionals)
are modeled and used to capture transitive dependences.
For each hierarchical level, we build the dependence relations $R_{\mathrm{dep}}$ over the gates visible at that level using
the Integer Set Library \cite{isl}. We capture ordering constraints among gates, and then compute its transitive closure,
{\footnotesize
$
R_{\mathrm{dep}}^{+} = \textsc{transitive\_closure}(R_{\mathrm{dep}}),
$}
and derive the dependence weight of a gate $g$ identified by $t$ as: 
$
\omega_g =
\left|
\{\, t' \mid [t] \!\to\! [t'] \in R_{\mathrm{dep}}^{+} \,\}
\right|.
$
Intuitively, $\omega_g$ measures how many gates (directly or indirectly) depend on $g$, capturing its global impact on the circuit. Gates with lower weights are less constrained by future computation and can be scheduled earlier with minimal disruption, while high-weight gates lie on many dependence paths and are more critical to the overall schedule \cite{qlosure}. Using the transitive closure ensures that $\omega_g$ reflects not only immediate dependences but all downstream effects, providing a global view that can be computed once and queried efficiently during analysis.

These abstractions form 
a robust foundation for analysis (Sec.~\ref{subsec:cost_function}) 
that enables us
to overcome limitations of DAG-based methods
on hierarchical, dynamic circuits.


\subsection{Inter-Circuit Qubit Remapping: Qubit Reconciliation}\label{subsec:reconciliation}
This pass computes a minimal sequence of SWAP operations that transforms
a {given placement} $\varphi_{\mathrm{init}}$ of one region into {another placement} $\varphi_{\mathrm{target}}$ required by the next.
Reconciliation is needed to revert back to a steady qubit state and for merging possible qubit states resulting
from sub-circuits of disjoint control-flow paths.
In practice, $\varphi_{init}$ is the final 
mapping state of a loop body, $\varphi_{body}$, or
the final state of a branch sub-circuit, $\varphi^{B_i}_{out}$ .

{\bf Formal Problem Definition.}
Given $G=(Q_{\mathrm{phys}}, R_{\mathrm{hw}})$, 
this problem 
computes
a sequence of remapping SWAPs $S$, that converts
$\varphi_{init}$ to $\varphi_{target}$.

This optimization problem is equivalent to the well-known
\emph{Token Swapping Problem}~\cite{bonnet2018complexity}, 
where each logical qubit corresponds to a token initially placed on a vertex
$\varphi_{\mathrm{init}}(q)$ of the hardware graph $G$, and the objective is
to move each token to its designated vertex $\varphi_{\mathrm{target}}(q)$
using the fewest possible adjacent swaps along $R_{hw}$.

\begin{figure}[t]
    \centering
    \includegraphics[width=0.8\linewidth]{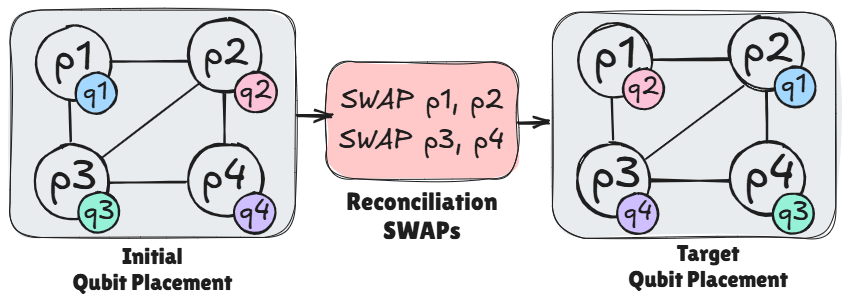}
    \shrink[4]
    \caption{
Token swapping between two qubit mappings, 
$\varphi_{\mathrm{init}}$ (left) and $\varphi_{\mathrm{target}}$. 
Physical qubits $p_i$ form the hardware graph, and logical qubits $q_i$ 
are labeled tokens. A sequence of SWAPs transforms the initial mapping 
into the target.
    }
    \Description{
    Token swapping between two qubit mappings, 
$\varphi_{\mathrm{init}}$ (left) and $\varphi_{\mathrm{target}}$. 
Physical qubits $p_i$ form the hardware graph, and logical qubits $q_i$ 
are labeled tokens. A sequence of SWAPs transforms the initial mapping 
into the target.
    }
    \label{fig:qroqi_reconciliation_example}
    \vspace{1ex}
\end{figure}

\paragraph*{Example.}
Fig.~\ref{fig:qroqi_reconciliation_example} depicts 
the mapping of four qubits ($q_{i}$) to four physical
locations, the $p_i$.
Links $q_l \rightarrow q_k$ denote usable pairs of qubits.
The logical-to-physical mappings before and after reconciliation are:

\noindent
\begin{minipage}{\linewidth}
{
\setlength{\abovedisplayskip}{2pt}
 \setlength{\belowdisplayskip}{2pt}
\FormulaSize
\[
\begin{aligned}
\varphi_{\mathrm{init}} &=
\{ q_1\!\to\!p_1,\; q_2\!\to\!p_2,\; q_3\!\to\!p_3,\; q_4\!\to\!p_4 \}, \\
\varphi_{\mathrm{target}} &=
\{ q_1\!\to\!p_2,\; q_2\!\to\!p_1,\; q_3\!\to\!p_4,\; q_4\!\to\!p_3 \}.
\end{aligned}
\]
}
\end{minipage}

\noindent
The reconciliation sequence 
$
S = [\mathrm{swap}(p_1, p_2),\, \mathrm{swap}(p_3, p_4)]$
transforms $\varphi_{\mathrm{init}}$ into $\varphi_{\mathrm{target}}
$,
yielding a consistent mapping for the next circuit block.

{\bf Reconciliation Methodology.}
As the remapping problem
is NP-hard,
\ourtool{} employs the 4-approximation algorithm proposed by
Miltzow et~al.~\cite{miltzow2016approximation}.
This method iteratively selects SWAPs on adjacent qubits that reduce their
aggregate distance to the target placement in the hardware connectivity graph,
producing a sequence that closely approximates the minimal number of swaps.
The algorithm runs in polynomial time and produces near-optimal reconciliation
sequences in practice, making it well suited for integration within
\ourtool{}’s compilation pipeline.

\new{Reconciliation enforces a target placement but does not choose it. The target depends
on the type of dynamic block: for a loop it is the steady-state entry mapping, and for a
conditional it is the mapping the branches merge into. How each is selected is described
in Sec.~\ref{subsec:routing_algorithm}.}


\subsection{Multi-Objective Mapping Cost Function 
}
\label{subsec:cost_function}
\ourtool{} maps straight-line sub-circuits using a heuristic derived from \textsc{Qlosure}~\cite{qlosure} which prioritizes SWAPs that reduce the dependency-weighted physical distance between dependent qubits. While effective for reducing routing overhead, this objective alone does not explicitly capture hardware error rates or the immediate impact of SWAPs on circuit depth and latency. To address this, \ourtool{} augments this heuristic with two complementary terms: a \emph{weighted-distance term} $C_{\mathrm{spatial}}^{(s)}$, which incorporates error rates into the distance metric, and a \emph{depth-rate term} $C_{\mathrm{rate}}^{(s)}$, which penalizes SWAPs on qubits already heavily used in the current mapping.

\begin{figure}[htb]
    \centering
    \begin{subfigure}[b]{0.48\linewidth}
        \centering
        \includegraphics[width=1\linewidth,height=2cm]{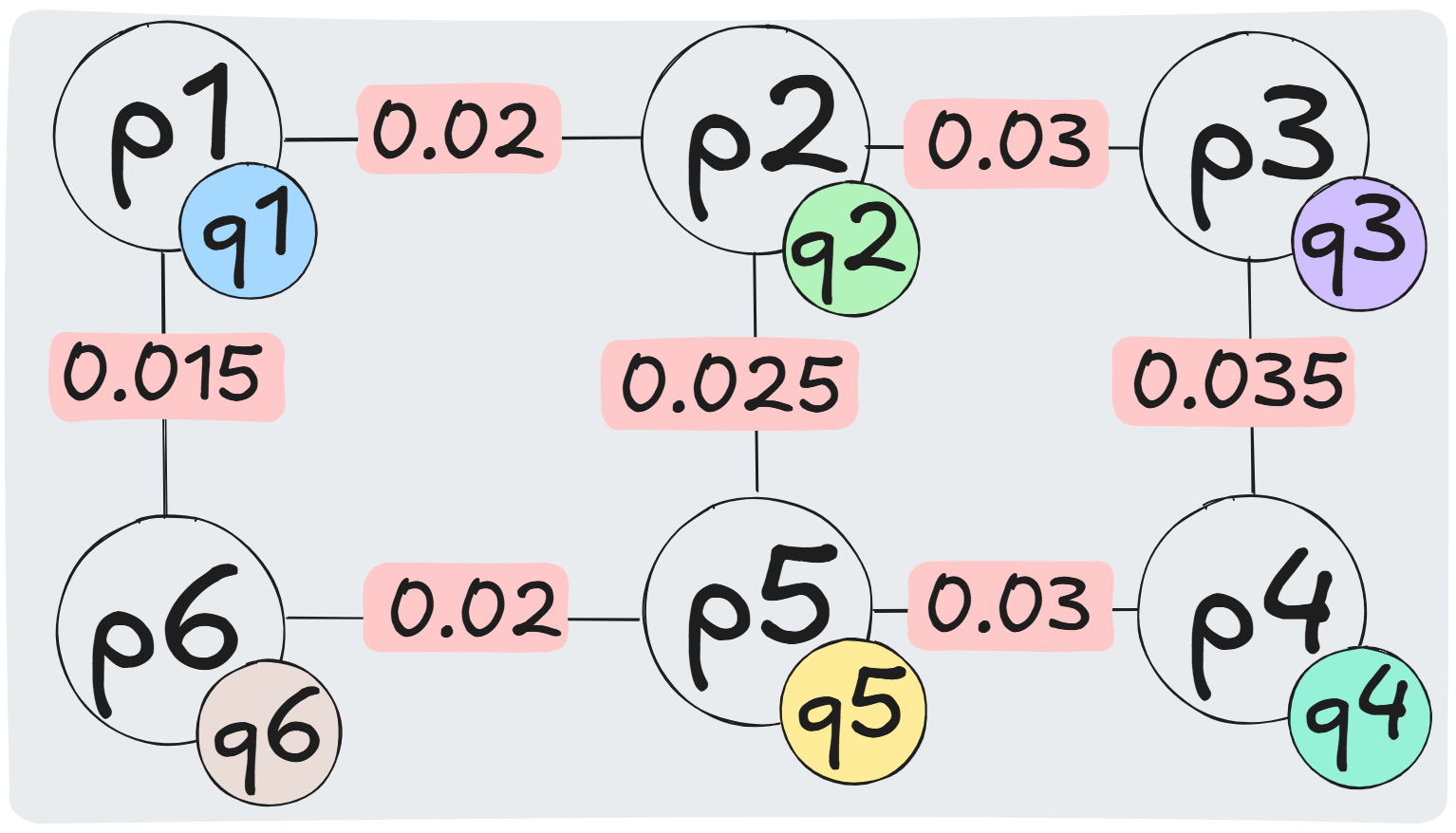}
        \vspace{-3ex}
        \caption{\footnotesize QPU topology 
        with  error rates.}
        \label{fig:hardware_topology}
    \end{subfigure}
    \hspace{.5em}
    \begin{subfigure}[b]{0.48\linewidth}
        \centering
        \includegraphics[width=1\linewidth,height=2cm]{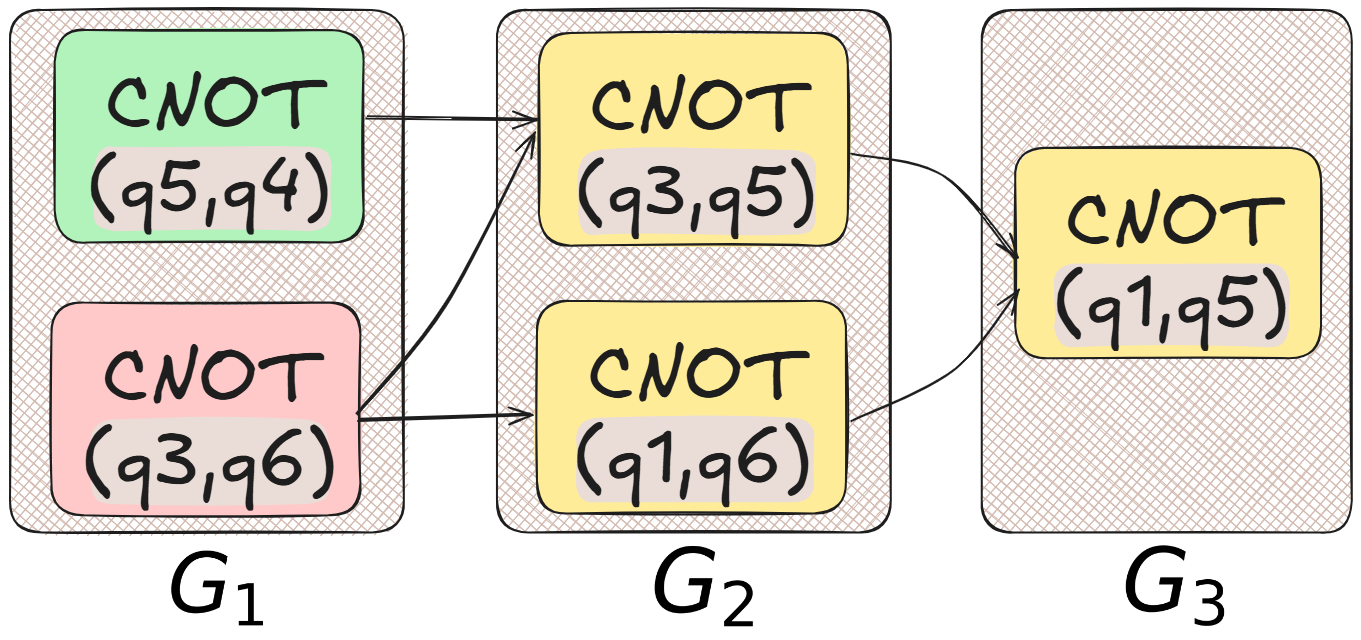}
        \vspace{-3ex}
        \caption{\footnotesize Circuit DAG view}
        \label{fig:logical_dependency}
    \end{subfigure}

    \shrink[3]
    \caption{Illustration of the physical topology and the corresponding logical gate dependencies used in the latency-aware dependence-driven mapping example. 
    Subfigure~(a) shows the 6-qubit hardware graph with coupling errors, while subfigure~(b) presents a logical circuit DAG.
    } 
    \label{fig:latency_aware_example}
    \Description{}
    \vspace{1ex}
\end{figure}

{\bf Weighted-Distance Term ($C^{(s)}_\mathrm{spatial}$).}
This term (Eq.~\ref{eq:D}) favors SWAPs that bring
interacting qubits closer while still avoiding high-error routing paths. 
Candidate SWAPs are evaluated over a
layered view of the circuit DAG (Fig.~\ref{fig:latency_aware_example}b), 
\new{
restricted to a lookahead window of 
$L$ dependence layers. The window is built from the current front layer — the set of gates whose dependences have all been satisfied — by collecting gates in topological order of the DAG and grouping them into layers such that all gates within a layer are mutually independent. It terminates once $L$ layers have been collected or the DAG is exhausted.
} 
For
each circuit layer $l$, we define $\Gamma_l^{(s)}$ as the dependency-weighted
average physical distance among interacting qubits after applying SWAP {\bf $s$} \cite{qlosure}. 
Then, we scale up
this distance by the average two-qubit gate error rate along
the corresponding shortest path in the hardware graph
(Fig.~\ref{fig:latency_aware_example}a), thereby explicitly incorporating
hardware reliability into the cost. 

\noindent
\begin{minipage}{\linewidth}
\begin{equation}
C_{\mathrm{spatial}}^{(s)} =
\frac{\sum_{l=1}^{L} w_l\, \Gamma_l^{(s)}}{\sum_{l=1}^{L} w_l},
\qquad
w_l = \frac{1}{l},
\label{eq:D}
\end{equation}
\end{minipage}

\noindent
Above, each layer $l$ is assigned a weight $w_l = 1/l$, so earlier layers (smaller $l$) contribute more to the score than later ones.

{\bf Depth-Rate Term ($C^{(s)}_{rate}$).}
Distance alone may still choose SWAPs that increase circuit depth. To reduce this effect, \ourtool{} adds a penalty for using qubits that already appear frequently in the current schedule. In Fig.~\ref{fig:latency_aware_example}, after executing CNOT$(q_4,q_5)$, swapping either participating qubit is more likely to increase the critical-path depth than using a less-active qubit. We capture this as

\noindent
\begin{minipage}{\linewidth}
\begin{equation}
C_{\mathrm{rate}}^{(s)}
=
\frac{
\max(\text{depth}(\varphi_s(s_{q_1})),\,\text{depth}(\varphi_s(s_{q_2})))
}{
\max_{p_i \in Q_{phys}}\text{depth}(p_i)
},
\label{eq:depth_rate}
\end{equation}
\end{minipage}

where $\text{depth}(p)$ is the number of executed gate layers involving physical qubit $p$. A high $C_{\mathrm{rate}}^{(s)}$ indicates that the SWAP uses qubits already on the critical path and is therefore penalized. This encourages the mapper to use other qubits when possible, improving parallelism and reducing latency.

\new{
{\bf In-Window Macro-nodes.}
These are built from
the \emph{current} control-flow level and region.
Macro-nodes may contain control-flow regions and ordinary gates. Such a region
contributes to $\Gamma^{(s)}_{l}$ in the same way any other node does: it is
treated as a single multi-qubit gate over the union of accessed logical qubits 
(Sec.~\ref{sec:overview}), so its pairwise qubit distances enter the
score directly, while its dependence weight $\omega_{g}$, derived from the
transitive closure $R^{+}_{\mathrm{dep}}$, already accounts
for its downstream impact on the rest of the circuit. Its \emph{internal} routing
is not evaluated here: it is resolved by the recursive call under the region's
own fixed $\varphi_{\mathrm{in}}/\varphi_{\mathrm{out}}$ mappings.
}

{\bf Composite Cost Function.}
Each candidate SWAP $s$ is scored as

\noindent
\begin{minipage}{\linewidth}
\begin{equation}
M(s) =
\max(\delta_{s_{p_1}}, \delta_{s_{p_2}})
\left[
\alpha\, C_{\mathrm{spatial}}^{(s)} +
\beta\, C_{\mathrm{rate}}^{(s)}
\right],
\label{eq:M_s}
\end{equation}
\end{minipage}

where $\delta$ is the standard decay factor used to discourage repeatedly swapping the same qubits~\cite{qlosure,sabre}. Parameters $\alpha$ and $\beta$ control the trade-off between reducing routing overhead and limiting circuit depth. In summary, $C_{\mathrm{spatial}}^{(s)}$ reduces communication cost and avoids high-error paths, while $C_{\mathrm{rate}}^{(s)}$ helps reduce circuit depth and latency.%

\subsection{Routing Algorithm}\label{subsec:routing_algorithm}

\ourtool{}'s
mapping 
relies on three
subcases 
(Alg.~\ref{alg:qroqi-loops} for loops, 
Alg.~\ref{alg:qroqi-bidir} for entry-mapping optimization, and
Alg.~\ref{alg:qroqi-conditionals} for conditionals),
and one driver algorithm,
Alg.~\ref{alg:routing:main}.
Sub-circuits are extracted from the bodies of
iterative and conditional
constructs, converted to 
a dual DAG-based and polyhedral-based representation, and later used to perform 
the necessary mapping.
To maintain mapping consistency, we assume that sub-circuits have single entry and exit points.

\begin{algorithm}[thb]
  \FormulaSize
  \caption{\footnotesize \ourtool{} — Handling \texttt{for}/\texttt{while} Loops
  }
  \label{alg:qroqi-loops}
  \begin{minipage}{\linewidth}
  
  \begin{algorithmic}[1]
    \Require Loop node $B$ with body $G_{\mathrm{body}}$; hardware map $R_{\mathrm{hw}}$; distance matrix $D_{\mathrm{phys}}$; current mapping $\varphi$ 
    \Ensure Complete compiled block  $C_{\text{entry}} \,\|\,$ $C_{\mathrm{loop}}$ 
    and updated mapping $\phistar{}$
    \State $\varphi_{\mathrm{in}}^\star \gets \textsc{SelectSteadyState}(G_{\mathrm{body}}, R_{\mathrm{hw}}, D_{\mathrm{phys}}, \varphi)$ \Comment{Alg.~\ref{alg:qroqi-bidir}}
    \State $\mathcal{S}_{\text{entry}} \gets \textsc{TokenSwap}(\varphi,\ \varphi_{\mathrm{in}}^\star,\ R_{\mathrm{hw}})$ \Comment{Reconciliation SWAPs}
    \State $C_{\text{entry}} \gets \textsc{EmitSwapSeq}(\mathcal{S}_{\text{entry}})$;\; $\varphi \gets \varphi_{\mathrm{in}}^\star$ \Comment{emits a sequence of SWAPs as IR}
    \State $[C_{\text{body}},\ \varphi_{\text{body}}] \gets \textsc{\ourtool{}}(G_{\mathrm{body}}, R_{\mathrm{hw}}, D_{\mathrm{phys}}, \varphi_{\mathrm{in}}^\star)$
    \State $\mathcal{S}_{\text{iter}} \gets \textsc{TokenSwap}(\varphi_{\text{body}},\ \varphi_{\mathrm{in}}^\star,\ R_{\mathrm{hw}})$  
    \State $C_{\mathrm{loop}} \gets \textsc{EmitLoop}\big(C_{\text{body}} \,\|\, \textsc{EmitSwapSeq}(\mathcal{S}_{\text{iter}})\big)$ 
    \State \Return $\big(C_{\text{entry}} \,\|\, C_{\mathrm{loop}},\ \varphi_{\mathrm{in}}^\star\big)$
  \end{algorithmic}
\end{minipage}
\end{algorithm}

\begin{algorithm}[thb]
  \FormulaSize
  \caption{\footnotesize SelectSteadyState: bidirectional passes 
  }
  \label{alg:qroqi-bidir}
  \begin{algorithmic}[1]
    \Require Loop body DAG $G_{\text{body}}$, hardware map $R_{\mathrm{hw}}$,
             distance matrix $D_{\mathrm{phys}}$, current mapping $\varphi$,
             number of passes $K\!\ge\!1$
    \Ensure Optimized loop entry mapping $\varphi_{\mathrm{in}}^\star$
    \State $\varphi_{\text{seed}}\gets \varphi$
    \For{$k=1$ to $K$}
      \State $\bigl[\_,\ \varphi_{\text{fwd\_out}}\bigr]\gets
             \textsc{\ourtool{}}(G_{\text{body}},R_{\mathrm{hw}},D_{\mathrm{phys}},\varphi_{\text{seed}})$
      \State $G_{\text{rev}}\gets \textsc{ReverseDAG}(G_{\text{body}})$ \Comment{invert edges and two-qubit direction 
      }
      \State $\bigl[\_,\ \varphi_{\text{bwd\_out}}\bigr]\gets
             \textsc{\ourtool{}}(G_{\text{rev}},R_{\mathrm{hw}},D_{\mathrm{phys}},\varphi_{\text{fwd\_out}})$
      \State $\varphi_{\text{seed}}\gets \varphi_{\text{bwd\_out}}$ \Comment{seed next round}
    \EndFor
    \State $\varphi_{\mathrm{in}}^\star \gets \varphi_{\text{seed}}$; \Return $\varphi_{\mathrm{in}}^\star$
    \Comment{final mapping after $K$ forward/backward passes}
  \end{algorithmic}
\end{algorithm} 



{\bf Mapping Iterative Constructs.} Alg.~\ref{alg:qroqi-loops} is used by \ourtool{} to
map iterative
control-flow structures, {\tt for-loops} and {\tt while-loops}.
First, a {\em steady-state} qubit mapping $\phistar$
is computed by the \textsc{SelectSteadyState} 
(Alg.~\ref{alg:qroqi-bidir}), which identifies a mapping that remains stable across loop iterations.
Once the steady-state
is decided, we invoke the \textsc{TokenSwap} algorithm \cite{miltzow2016approximation} to determine the
sequence of SWAP gates that convert the initial
mapping state, $\varphi_{in}$—i.e., the mapping at the point just before entering the loop—into $\phistar$.
The found sequence of SWAP gates, 
$\mathcal{S}_{\text{entry}}$, is emitted as a sequence of SWAP operations and prepended to the loop body circuit $C_{\text{body}}$ (i.e., the compiled representation of the loop body), after which the current mapping state is updated to $\varphi_{\mathrm{in}}^\star$.

\begin{figure*}[thb]
    \centering
    \begin{subfigure}[b]{0.49\textwidth}
        \includegraphics[width=\linewidth]{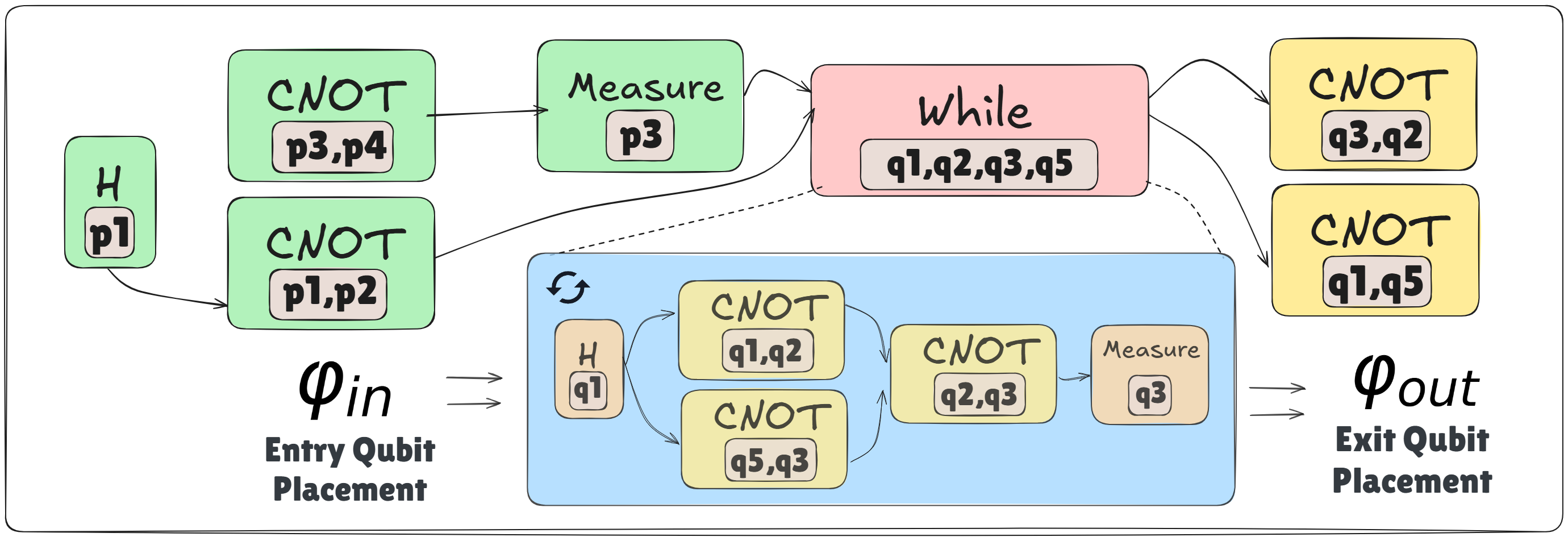}
        \vspace{-4ex}
        \caption{DAG representation of a circuit.}
        \label{fig:dag-repr}
    \end{subfigure}
    \begin{subfigure}[b]{0.49\textwidth}
        \includegraphics[width=\linewidth]{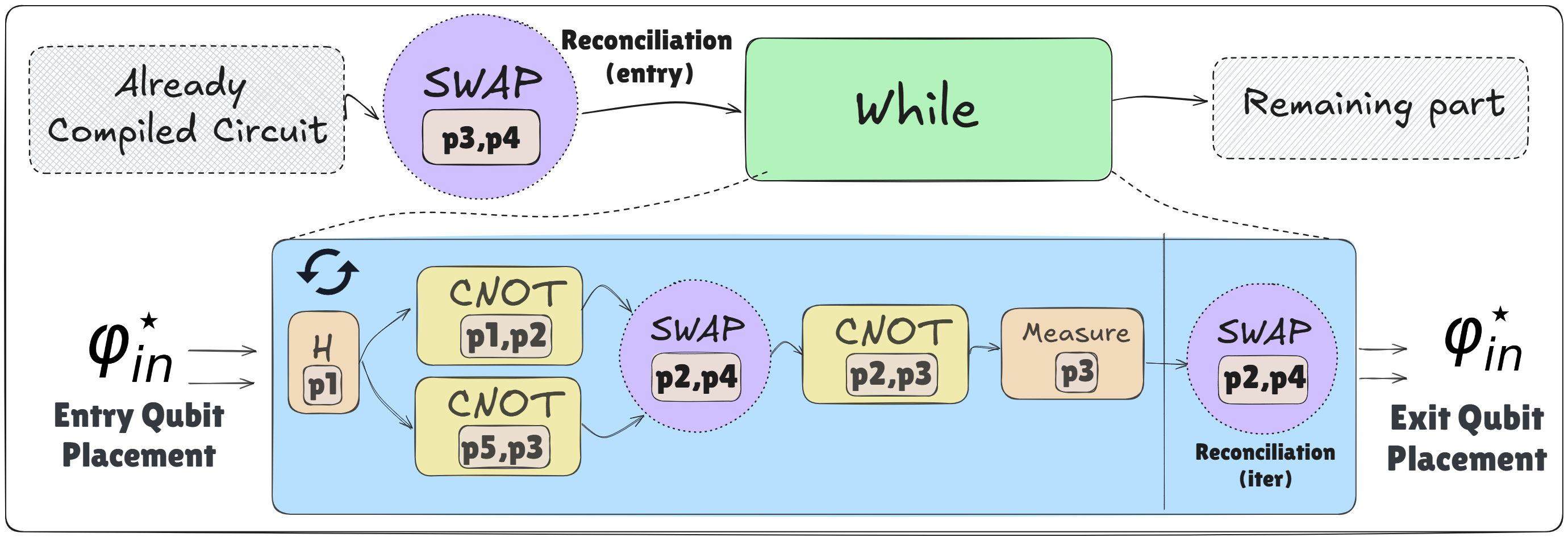}
        \vspace{-4ex}
        \caption{Compilation and reconciliation process.
        }
        \label{fig:compilation-recon}
    \end{subfigure}
       \vspace{-2ex}
    \caption{
        Illustration of the loop compilation process performed by \ourtool{}. 
        Subfigure~(a) shows the DAG-level structure of the circuit with 
        qubit placements $\varphi_{\mathrm{in}}$ and $\varphi_{\mathrm{out}}$. 
        Subfigure~(b) depicts the compilation flow integrating two reconciliation 
        passes: one at loop entry (\mbox{$\mathcal{S}_{\text{entry}}$}) 
        to align the qubit layout with the steady-state mapping 
        $\varphi_{\mathrm{in}}^\star$, and one per iteration 
        (\mbox{$\mathcal{S}_{\text{iter}}$}) to restore this mapping after each 
        iteration. These transformations respect the device connectivity 
        constraints shown in Fig.~\ref{fig:hardware_topology}.
    }
    \vspace{-1ex}
    \label{fig:loop-reconciliation}
    \Description{}
\end{figure*}

As illustrated in Fig.~\ref{fig:loop-reconciliation}, \ourtool{} then takes the loop body $B_{body}$ and maps
individual sub-circuits, those in disjoint control-flow paths, 
in a recursive manner, returning a mapped circuit body $C_{body}$ and the outgoing qubit state from it, $\varphi_{out}$.
The latter is reconciled back to $\phistar$ via \textsc{TokenSwap}, generating a sequence $\mathcal{S}_{\text{iter}}$ that ensures consistency across iterations.
Finally, these reconciliation SWAPs are appended to the loop body, forming a self-consistent compiled loop.

As having 
input/output mappings that require few SWAPs
is necessary, we use 
Alg.~\ref{alg:qroqi-bidir} to search 
for such mapping states via
a short-fixed number of passes $K$, using bidirectional passes:
alternating forward and backward executions of the mapping algorithm over the circuit and its reverse, repeated $K$ times, typically converging after 3--4 passes.
A similar approach
has been previously used in \cite{sabre} to find
initial logical-to-physical qubit mappings, using
$K=3$.

\begin{algorithm}[H]
  \FormulaSize
  \caption{\footnotesize \ourtool{} — Handling \texttt{if-else} Conditionals}
  \label{alg:qroqi-conditionals}
  \begin{algorithmic}[1]
    \Require Conditional node $B$ with branch subgraphs $\{G_i\}$; hardware map $R_{\mathrm{hw}}$; distance matrix $D_{\mathrm{phys}}$; entry mapping $\varphi$
    \Ensure 
    reconciled conditional $C_{\mathrm{cond}}$ and updated mapping $\varphi$
    \For{each branch $i$}
    \label{alg:groqi:branches:mapping:start}
      \State $[C_i,\ \varphi^{B_i}_{\text{out}}] \gets \textsc{\ourtool{}}(G_i, R_{\mathrm{hw}}, D_{\mathrm{phys}}, \varphi)$
    \EndFor
    \label{alg:groqi:branches:mapping:finish}
    \State $\varphi_{\mathrm{merge}} \gets \textsc{ChooseUnifiedMapping}\big(\{\varphi^{B_i}_{\text{out}}\}\big);\;$ $C_{\mathrm{cond}} \gets \textsc{BeginIfElse}()$
    \label{alg:groqi:branches:recon:start}
    \For{each branch $i$}
      \State $\mathcal{S}_i \gets \textsc{TokenSwap}(\varphi^{B_i}_{\text{out}},\ \varphi_{\mathrm{merge}},\ R_{\mathrm{hw}})$
      \State $C_{\mathrm{cond}} \gets C_{\mathrm{cond}} \,\|\, \textsc{EmitBranch}(C_i \,\|\, \textsc{EmitSwapSeq}(\mathcal{S}_i))$
    \EndFor
    \label{alg:groqi:branches:recon:finish}
    \State $C_{\mathrm{cond}} \gets \textsc{EndIfElse}(C_{\mathrm{cond}})$;\; $\varphi \gets \varphi_{\mathrm{merge}};\;$ \Return $\big(C_{\mathrm{cond}},\ \varphi\big)$
  \end{algorithmic}
\end{algorithm}

{\bf Mapping Conditional Constructs.}
Alg.~\ref{alg:qroqi-conditionals} presents
the steps to map sub-circuits in disjoint control-flow paths (lines \ref{alg:groqi:branches:mapping:start}--\ref{alg:groqi:branches:mapping:finish}).
Each sub-circuit $G_i$ in each path $i$ is mapped
with 
Alg.~\ref{alg:routing:main},
\old{
Then, the resulting branch with the
lowest cost
is selected as
$\varphi_{merge}$.
After this, we find
and append
reconciliation passes that convert each
$\varphi^{B_i}_{out}$ to
$\varphi_{merge}$.
}
\new{
yielding one exit
mapping $\varphi^{B_{i}}_{out}$ per branch. \textsc{ChooseUnifiedMapping} (Alg.~\ref{alg:qroqi-conditionals},
line~\ref{alg:groqi:branches:recon:start}) then merges these into a single
placement: it scores each
$\varphi^{B_{i}}_{out}$ with $C_{\mathrm{spatial}}$ (Eq.~\ref{eq:D}),
evaluated over the lookahead window built from the front layer immediately
\emph{after} the merge point, and keeps the 
lowest-scoring mapping as $\varphi_{merge}$. The
score therefore weighs downstream routing overhead, gate criticality through
$\omega_{g}$, and hardware reliability through the per-edge
error rates. 
}

\new{
{\bf Branch Probabilities.}
\ourtool{} weights all branches with
equal probability, 
which is the default when the
branch distribution 
is unknown at compile-time. 
When probabilities $p_{i}$ are
available, they can be folded in by scaling each score by $(1-p_{i})$. All
experiments assume uniform probabilities.
}

{\bf Putting It All Together.} 
Finally, we present the  driver algorithm used
in \ourtool{}. The algorithm first performs 
initialization tasks (lines \ref{alg:groqi:main:init:start}--\ref{alg:groqi:main:init:end}) that create an empty circuit,
prepare the circuit front layer (with schedulable gates),
compute the transitive closure among gates
\cite{qlosure}, and start qubit decay factors (which avoid reusing the same qubit in mapping paths for a short time frame).
The main loop (lines \ref{alg:groqi:main:loop:start}--\ref{alg:groqi:main:loop:finish}) handles the three
distinct cases of a circuit: 
 i) when executable gates are ready (lines \ref{alg:dynamiq:main:ready-case:start}--\ref{alg:dynamiq:main:ready-case:end}), they are directly emitted and the front layer is advanced;
ii) when the front layer contains only control-flow constructs (lines 10--16), a block is selected (line \ref{alg:dynamiq:main:select-dynamic-bloc}) and handled via Alg.~\ref{alg:qroqi-loops} or Alg.~\ref{alg:qroqi-conditionals} (lines \ref{alg:dynamiq:main:handle-dynamic-bloc:start}--\ref{alg:dynamiq:main:handle-dynamic-bloc:end}), after which the compiled block is appended (line \ref{alg:dynamiq:main:append-dynamic-block}) and spliced out of the DAG (line \ref{alg:dynamiq:main:splice-dynamic-block});
iii) otherwise, SWAP candidates are generated, evaluated using the cost model, and the best SWAP is applied to update the mapping and circuit (lines \ref{alg:dynamiq:main:choose-swap:start}--\ref{alg:dynamiq:main:choose-swap:end}).

Once a sub-circuit $B$ exhibiting control-flow is mapped,
$B$ is spliced out of $G$ to prepare for subsequent gates.

\begin{algorithm}[tbh]
  \FormulaSize
  \caption{\footnotesize \ourtool{}: Recursive Routing over Dynamic Blocks
  }
  \label{alg:routing:main}
  \begin{minipage}{\linewidth}
  \begin{algorithmic}[1]
    \Require Outer DAG $G$ (gates, measurements, dynamic blocks), hardware map $R_{\mathrm{hw}}$, distance matrix $D_{\mathrm{phys}}$, initial mapping $\varphi$
    \Ensure Routed circuit $C$ and final mapping $\varphi$
    \State $C \gets \epsilon$ \Comment{empty output circuit}
    \label{alg:groqi:main:init:start}
    \State $L_{\mathrm f} \gets \textsc{InitFrontLayer}(G)$
    \State $R^{+} \gets \textsc{TransitiveClosureTwoQ}(G)$ \Comment{for scoring only}
    \State $\delta \gets \mathbf{1}$ on $Q_{\mathrm{phys}}$ \Comment{decay/usage vector}
    \label{alg:groqi:main:init:end}
    \While{$L_{\mathrm f} \neq \varnothing$}
    \label{alg:groqi:main:loop:start}
      \State $G_{\text{ready}} \gets \textsc{ExtractReadyGates}(L_{\mathrm f}, \varphi, R_{\mathrm{hw}})$
      \If{$G_{\text{ready}} \neq \varnothing$}
      \label{alg:dynamiq:main:ready-case:start}
        \State $C \gets C \,\|\, \textsc{EmitGates}(G_{\text{ready}})$
        \State \textsc{Advance}$(G_{\text{ready}}, G, L_{\mathrm f})$;\; $\delta \gets \mathbf{1}$
        \label{alg:dynamiq:main:ready-case:end}
      \ElsIf{\textsc{HasOnlyControlFlow}$(L_{\mathrm f})$} 
        \State $B \gets \textsc{PickControlNode}(L_{\mathrm f})$ \Comment{\textnormal{\texttt{for}/\texttt{while} or \texttt{if-else}}}
        \label{alg:dynamiq:main:select-dynamic-bloc}
        \If{$\text{type}(B) \in \{\texttt{for},\texttt{while}\}$}
        \label{alg:dynamiq:main:handle-dynamic-bloc:start}
          \State $[C_B,\ \varphi] \gets \textsc{HandleLoop}(B,\ R_{\mathrm{hw}},\ D_{\mathrm{phys}},\ \varphi)$ \Comment{Alg.~\ref{alg:qroqi-loops}}
        \Else 
          \State $[C_B,\ \varphi] \gets \textsc{HandleConditional}(B,\ R_{\mathrm{hw}},\ D_{\mathrm{phys}},\ \varphi)$ 
          \Comment{Alg.~\ref{alg:qroqi-conditionals}}
        \EndIf
        \label{alg:dynamiq:main:handle-dynamic-bloc:end}
        \State  $C \gets C \,\|\, C_B$;
        \label{alg:dynamiq:main:append-dynamic-block}
        \State \textsc{SpliceCompiledBlockAndAdvance}$(G,B,L_{\mathrm f})$
        \Comment{update $L_f$}
        \label{alg:dynamiq:main:splice-dynamic-block}
      \Else 
        \State
        $L_{\mathrm w} \gets \textsc{MakeLookahead}(L_{\mathrm f}, G)$
        \State
        $\mathcal{G} \gets \textsc{BuildLookaheadLayers}(L_{\mathrm w}, G)$
        \label{alg:dynamiq:main:choose-swap:start}
        \State $\mathcal{S} \gets \textsc{GenSwapCandidates}(L_{\mathrm f}, R_{\mathrm{hw}})$
        \State $s^\star \gets \underset{s \in \mathcal{S}}{\arg\min}\; M(s,\ R^{+},\ \mathcal{G},\ D_{\mathrm{phys}},\ \delta)$ \Comment{Sec.~\ref{subsec:cost_function}}
        \State $\varphi \gets \varphi \circ s^\star$;\; $C \gets C \,\|\, \textsc{EmitSwap}(s^\star)$;\; $\delta \gets \textsc{UpdateDecay}(\delta, s^\star)$
        \label{alg:dynamiq:main:choose-swap:end}
      \EndIf
    \EndWhile
    \label{alg:groqi:main:loop:finish}
    \State \Return $(C,\ \varphi)$
  \end{algorithmic}
\end{minipage}
\end{algorithm}


\noindent\textbf{Example
(Fig.~\ref{fig:loop-reconciliation}):}
Mapping starts by fixing connectivity for the static gates, 
adding and scheduling the necessary SWAPs where required. 
In Fig.~\ref{fig:loop-reconciliation}(a), 
the gates shown in green can all be executed directly 
because their corresponding physical qubits are adjacent 
on the target hardware topology (Fig.~\ref{fig:hardware_topology}). 
Next, we encounter a dynamic block—a {\tt while}-loop—and thus invoke Alg.~\ref{alg:qroqi-loops}. 
We begin by identifying the steady-state qubit mapping 
$\varphi_{in}^{\star}$ using Alg.~\ref{alg:qroqi-bidir} on the body of the loop. 
To reconcile the current mapping with this steady state, 
an additional SWAP between $p_3$ and $p_4$ is inserted before entering the loop, 
as shown in Fig.
~\ref{fig:compilation-recon}.
This operation tries to resolve certain disconnections inside the loop 
before actually entering it. 
Next, we recursively invoke \ourtool{} on the loop body, 
where another SWAP between $p_2$ and $p_4$ is introduced to enable execution of 
the {\tt CNOT}$(q_2,q_3)$ gate. 
Because this SWAP changes $\varphi_{in}^{\star}$, 
a final reconciliation pass is added at the end of the loop 
to restore $\varphi_{in}^{\star}$ for the next iteration. 
Once the loop finishes execution, compilation proceeds 
with the remaining portion of the circuit.
\section{Dynamic-QUEKO: Hierarchical Circuits}
\label{sec:queko-loops}

We build circuits from a simple grammar over node types:
{\footnotesize
$\mathcal{N} ::= \texttt{leaf} \mid \texttt{seq}(\mathcal{N}^+) \mid \texttt{for}(K,\mathcal{N}) \mid \texttt{while}(q,\mathcal{N}) \mid \texttt{ifelse}(q,\mathcal{N},\mathcal{N})$}.

A \texttt{leaf} is a static 
block with 
depth \texttt{leaf\_depth} and one- and two-qubit gate densities $(\delta_1,\delta_2)$
where $(\delta_1,\delta_2)$ specify the fraction of qubits active per 
layer.
\texttt{seq} composes subcircuits sequentially, \texttt{for}$(K,\mathcal{N})$ 
is a counting-loop with $K$ iterations, 
\texttt{while}$(q,\mathcal{N})$ 
is 
a measurement-controlled loop 
and finally \texttt{ifelse}$(q,\mathcal{N},\mathcal{N})$
defines a conditional branch controlled by a measurement outcome associated with qubit $q$. 
This grammar 
generates nested, parameterized circuits.
Examples
of generated 
circuits are included in the supplemental material.

d-\textsc{QUEKO}
circuits generate an underlying
topology for a
given number of physical qubits, \texttt{n\_qubits}.
Edges are added with probability $\rho \in [0,1]$. 
This 
grants
control over routing difficulty: 
$\rho$ sets the connectivity density (higher $\rho$ increases parallelism).
A density of $\rho = 1$ corresponds to a complete graph. 
$\rho \times (|V|-1)$ is the average number of links for any given qubit. 
\texttt{leaf\_depth}
sets the depth of each static region and is the main knob for 
complexity.


\section{Evaluation}
\label{sec:eval}
\subsection{Experimental Testbed}

We validate the efficiency of \ourtool{} 
against 
QISKIT
\new{
1.3.2
and its default routing pass,
LightSABRE \cite{zou2024lightsabre},
}
the only mapper that supports circuits
with dynamic control-flow.
\new{
LightSABRE is invoked with \texttt{routing\_method=} 'sabre' and \texttt{layout\_method=} 'identity', \texttt{optimization\_level=1}, all remaining options at their defaults, and \texttt{seed\_transpiler} swept over 10 seeds. We select this setup to preserve the gate count and the initial placement across both
mappers, and focus on the impact of control-flow in the mapping process. For the IBM QPUs, the 127-qubit Brisbane and the 156-qubit Kingston, we use
calibration data collected on October~17, 2025.}
Experiments aim to demonstrate the quality
of our mapper across four
metrics: circuit depth, SWAP count, circuit
latency, and incurred error along the critical path.

This evaluation consists of five experiments:
a) Benchmarks on {\tt d-QUEKO}
(Sec.~\ref{sec:exp:dynamic-synthetic}),
b) a comparison on chiplet architectures
(Sec.~\ref{sec:chiplet-results-section}),
c) an experiment using surface codes
(Sec.~\ref{sec:surface-code-results}),
d) scalability results (Sec.~\ref{sec:eval:scalability}),
and
e) ablation studies (Sec.~\ref{sec:exp:ablations}).

\new{
{\bf Evaluated Circuits.}
We use dynamic 
circuits from {\tt d-QUEKO} and 
{\tt Stim}~\cite{gidney2021stim}.
{\tt d-QUEKO}
exposes data-dependent control
flow with controlled coverage of the structural dimensions that define
mapping difficulty: nesting depth and control-flow shape (grammar $\mathcal{N}$),
static-region size (\texttt{leaf\_depth}, swept 10--90), circuit width
(54/81/121/256 qubits), and gate density $(\delta_1,\delta_2)$. 
With Stim
we generate 
rotated surface-code circuits (Sec.~7.4) 
with standard, externally-defined QEC workloads: memory-$Z$ experiments swept over code distances $d\in\{3,5,7\}$ and rounds $r\in\{3,5,10,15,20\}$, providing an independent check on {\tt d-QUEKO} results
in the regime of adaptive quantum circuits.
}

\paragraph*{Performance Metrics}
We use four metrics to compare quantum mappers:
\textbf{SWAP count} measures the total number of \texttt{SWAP} operations in the trace;
\textbf{Circuit depth} measures the minimal number of parallel layers under qubit exclusivity;
\textbf{Latency} accumulates per-qubit gate durations from hardware calibration
data, while \textbf{Error} accumulates per-gate physical error rates on each qubit and averages them to obtain the overall error of the compiled circuit,  similar to prior work \cite{siraichi2019qubit}. 
These metrics are extracted from the 
CP (Critical Path) of mapped circuits, except for the SWAP count, which
considers the total number of SWAP gates introduced by each mapper.
We assume that all loops execute for ten iterations. 
For conditional blocks, all branches are evaluated and the branch with the worst-case result is selected for every metric.
The same methodology is applied, recursively, across nested
constructs. The improvements ($\Delta$\%) reported in the tables are computed for each metric as 
$(\mathrm{Val}_{\text{baseline}} - \mathrm{Val}_{\text{ours}}) / \mathrm{Val}_{\text{baseline}} \times 100$.

{\bf Research Questions.} Our evaluation aims to answer the following research questions: RQ1) Is \ourtool{} more effective than baseline methods? RQ2) Are qubit reconciliation, loop entry remapping, and the multi-objective function 
the underlying reasons for the improvement in dynamic circuits?

\subsection{Evaluating \ourtool{} on the d-QUEKO Benchmark}
\label{sec:exp:dynamic-synthetic}

This section studies
the impact of \ourtool{} on {\tt d-QUEKO} dynamic circuits.
We generate
benchmarks of the form:  \\
{
\footnotesize
$
\mathcal{N}
= \texttt{seq}~\!\Big( (\texttt{leaf},\;\texttt{while}(q,\texttt{leaf}),\;\texttt{leaf}) ^{\times 10} \Big),
$
}
i.e., the three-block pattern is placed \emph{ten times} in a row.  
For each such structure we generate 10 distinct circuits,  
and we do this for circuits using \textbf{54}, \textbf{81}, and \textbf{121} qubits, while varying leaf depths from 10 to 90.
Later, in Sec.~\ref{sec:exp:dynamic:loop-depth}, we extend 
the exploration to deeper nested structures defined by \(\mathcal{N}\).





\begin{table}[H]
\TableFontSize
\caption{
Average relative improvements (\%) of \ourtool{} over Qiskit LightSABRE across two IBM backends (Higher is better).
Benchmark names are abbreviated as dQ-$n$, where $n$ is the qubit count. Metric abbreviations are:
S = SWAPs, D = Depth, L = Latency, and E = Error.
}
\label{tab:results_dynamic_avg_improvements}
\shrink[2]
\centering
\begin{tabular}{lcccc|cccc}
\toprule
 & \multicolumn{4}{c}{\textbf{Brisbane (127 qbt)}} 
 & \multicolumn{4}{c}{\textbf{Kingston (156 qbt)}} \\
\cmidrule(lr){2-5} \cmidrule(lr){6-9}
\textbf{Bench.} 
  & \textbf{S} & \textbf{D} & \textbf{L} & \textbf{E}
  & \textbf{S} & \textbf{D} & \textbf{L} & \textbf{E} \\
\midrule
dQ-54  & 52.04 & 13.92 & 13.06 & 14.62  & 39.46 & 11.48 & 11.16 & 40.53 \\
dQ-81  & 40.97 & 14.66 & 13.99 & 21.08  & 48.71 & 18.48 & 18.35 & 38.99 \\
dQ-121 & 51.73 & 17.75 & 18.59 & 16.87  & 34.54 &  8.70 &  8.80 & 33.62 \\
\bottomrule
\end{tabular}
\end{table}

\begin{figure*}[htbp]

\vspace{-1ex}
\centering

    \setlength{\tabcolsep}{1pt} 
    \captionsetup[subfigure]{skip=0pt}
    \resizebox{0.95\textwidth}{!}{ 
    \begin{tabular}{@{}cccc@{}}

        \subcaptionbox{\footnotesize 81qbt -- SWAPs}[0.24\textwidth]{%
            \includegraphics[width=\linewidth]{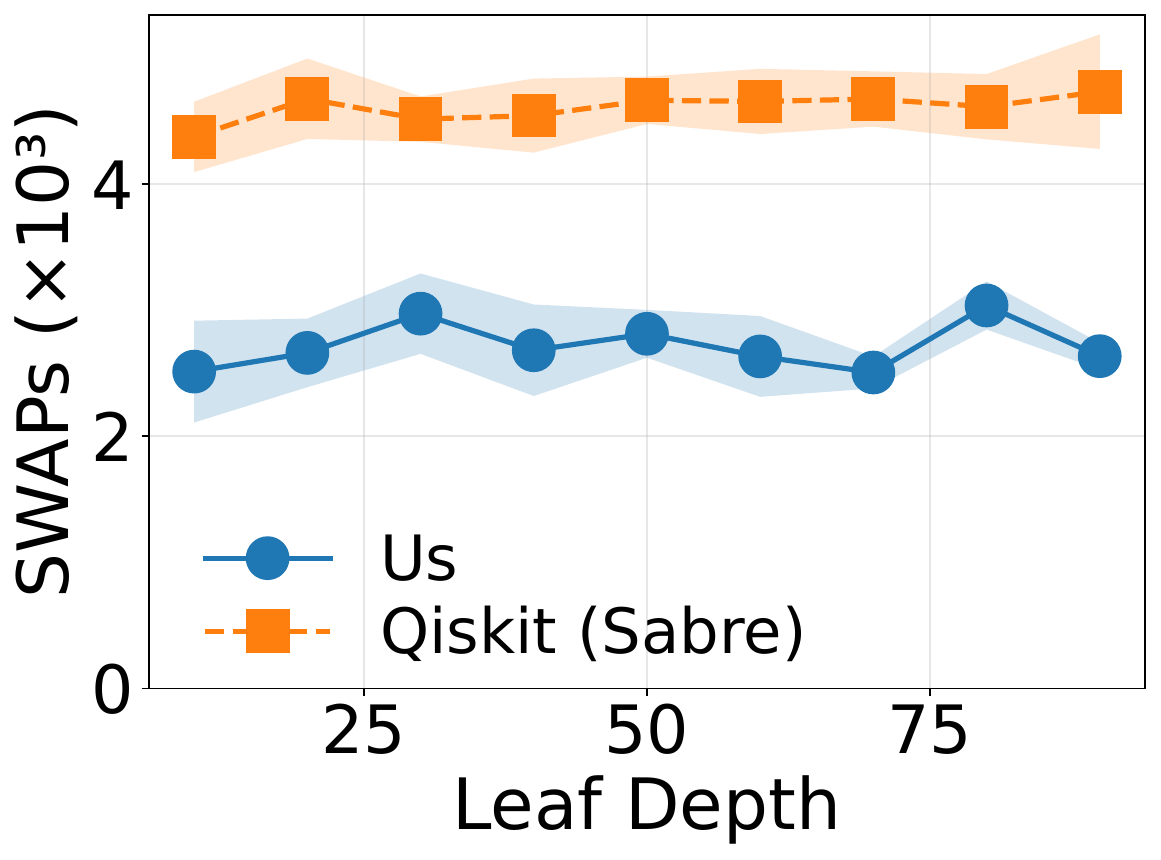}} &
        \subcaptionbox{\footnotesize 81qbt - Depth}[0.24\textwidth]{%
            \includegraphics[width=\linewidth]{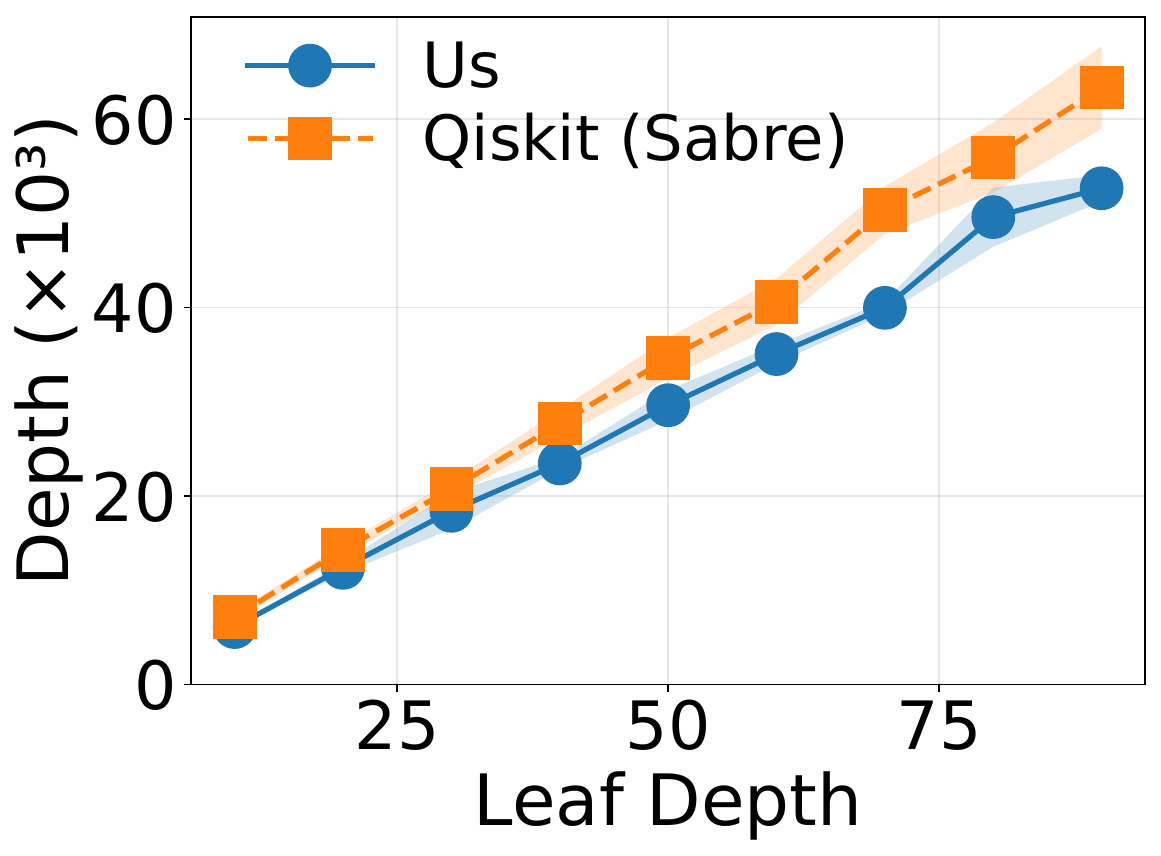}} &
        \subcaptionbox{\footnotesize 81qbt -- Latency}[0.24\textwidth]{%
            \includegraphics[width=\linewidth]{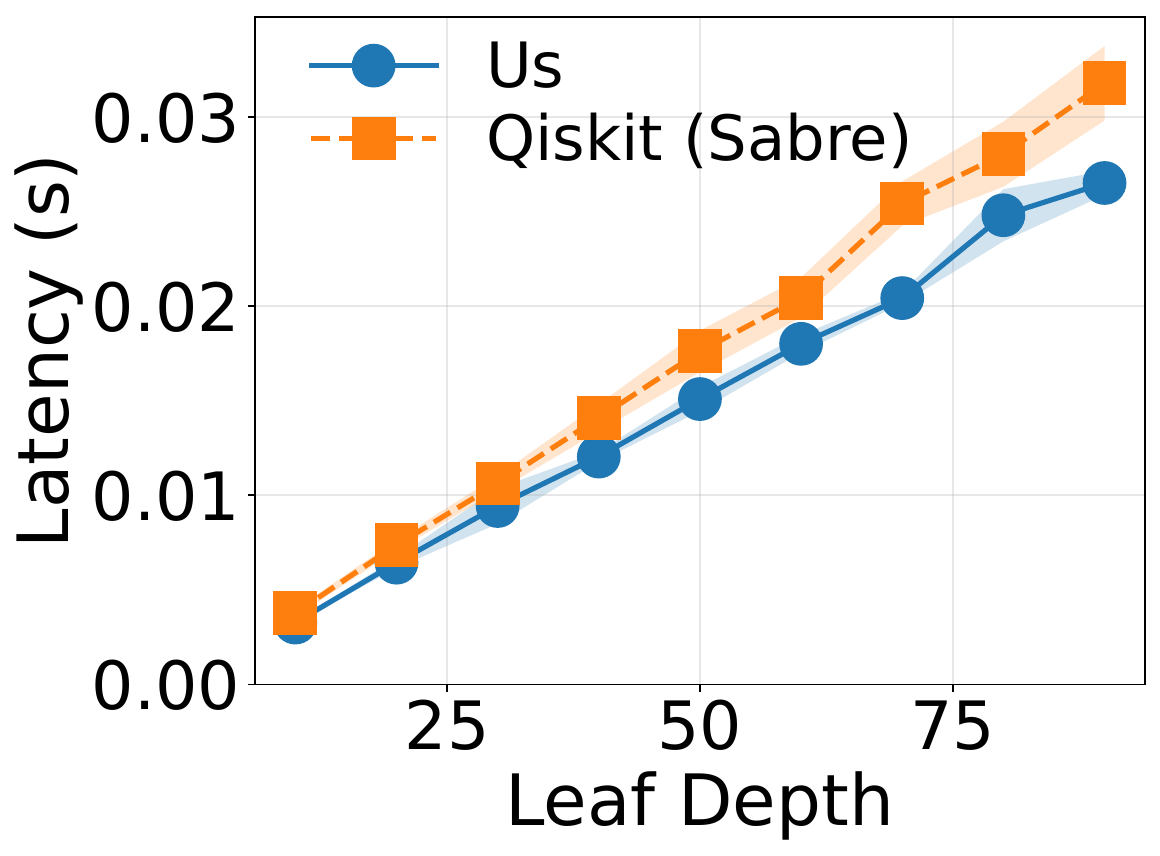}} &
        \subcaptionbox{\footnotesize 81qbt -- Error}[0.24\textwidth]{%
            \includegraphics[width=\linewidth]{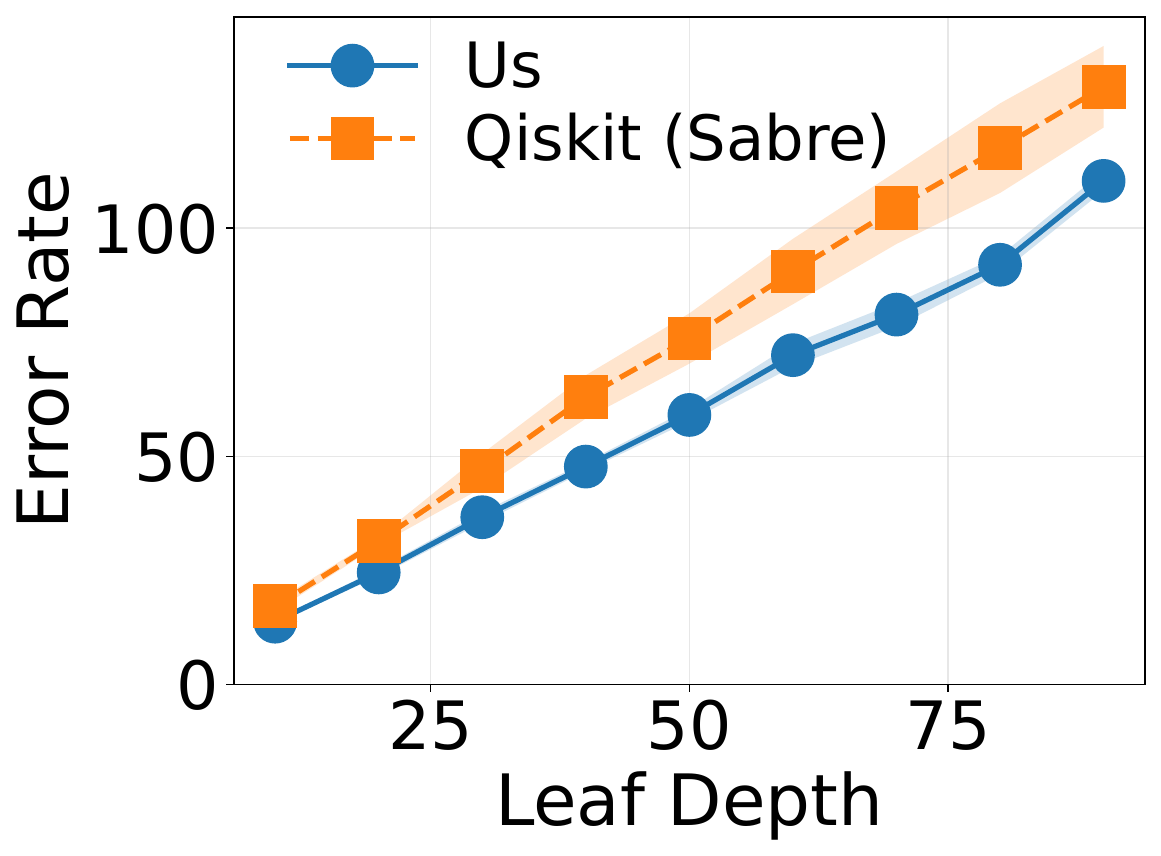}} \\[0.5em]

        \subcaptionbox{\footnotesize 121qbt -- SWAPs}[0.24\textwidth]{%
            \includegraphics[width=\linewidth]{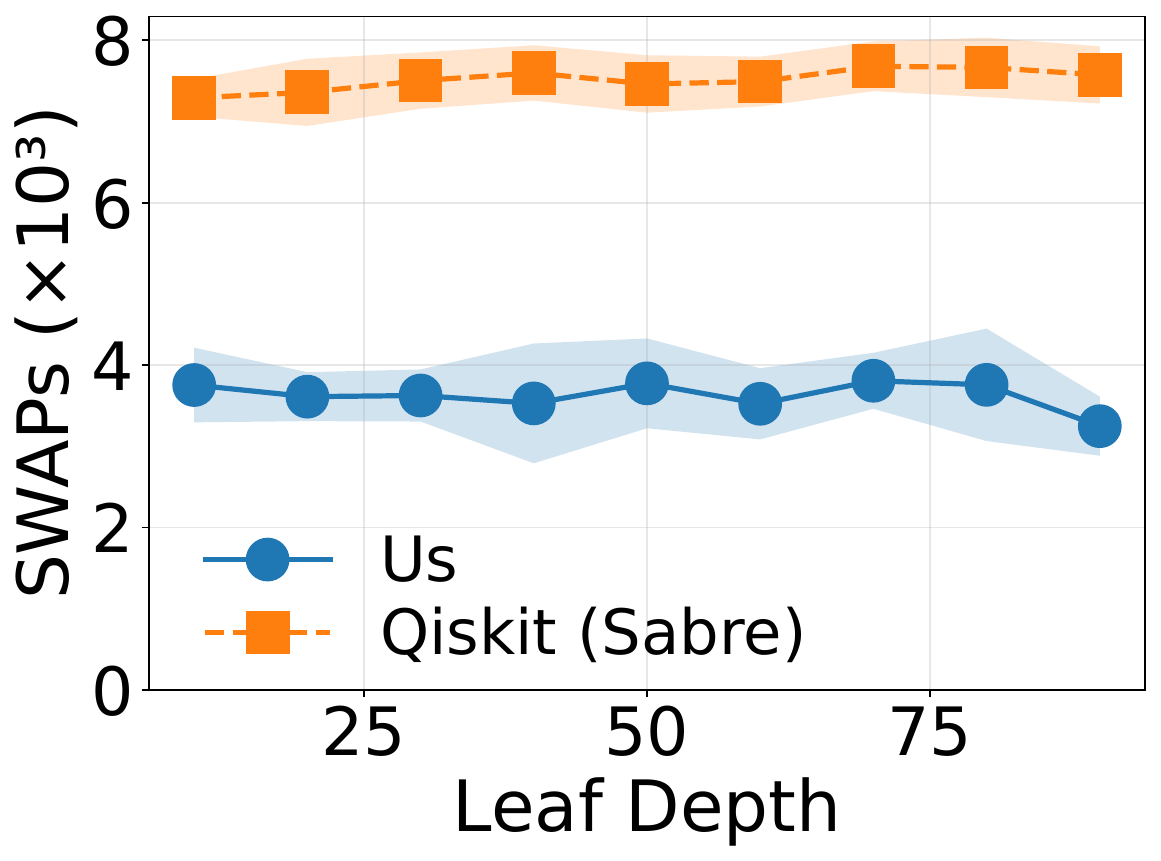}} &
        \subcaptionbox{\footnotesize 121qbt -- Depth}[0.24\textwidth]{%
            \includegraphics[width=\linewidth]{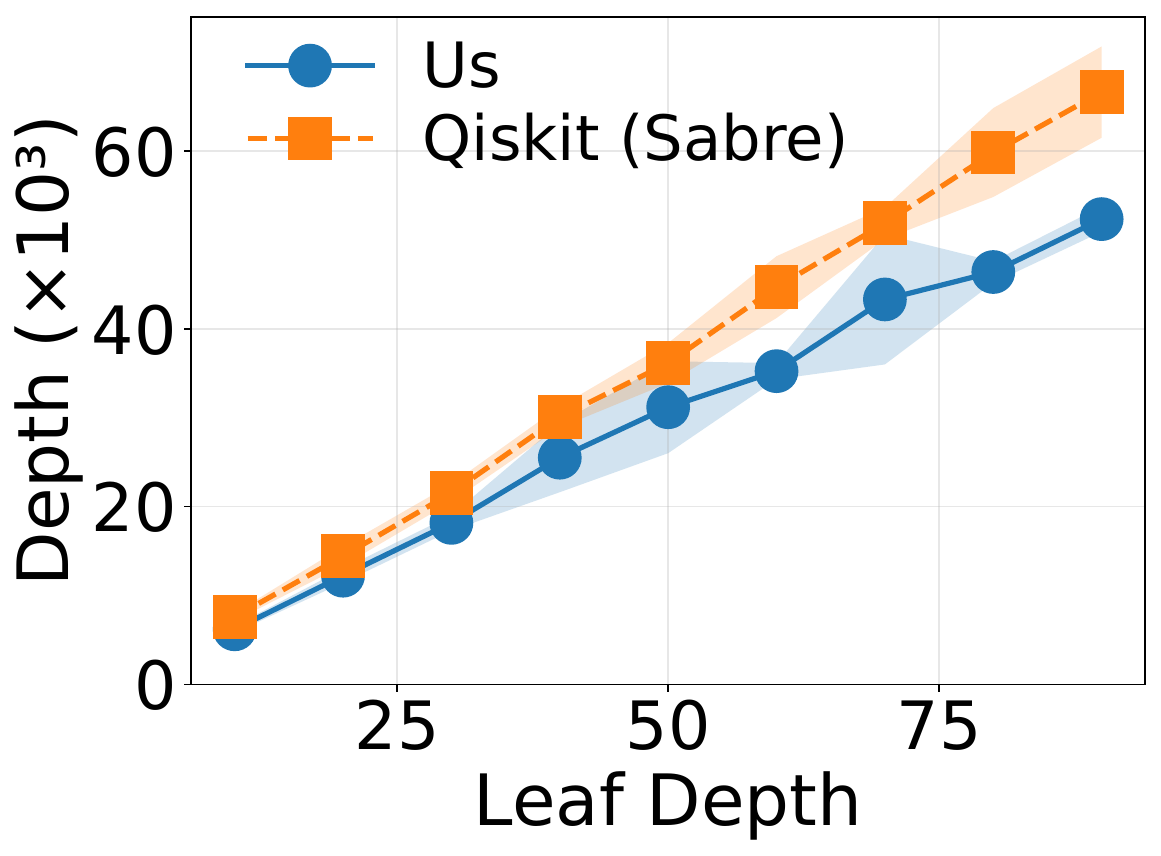}} &
        \subcaptionbox{\footnotesize 121qbt -- Latency}[0.24\textwidth]{%
            \includegraphics[width=\linewidth]{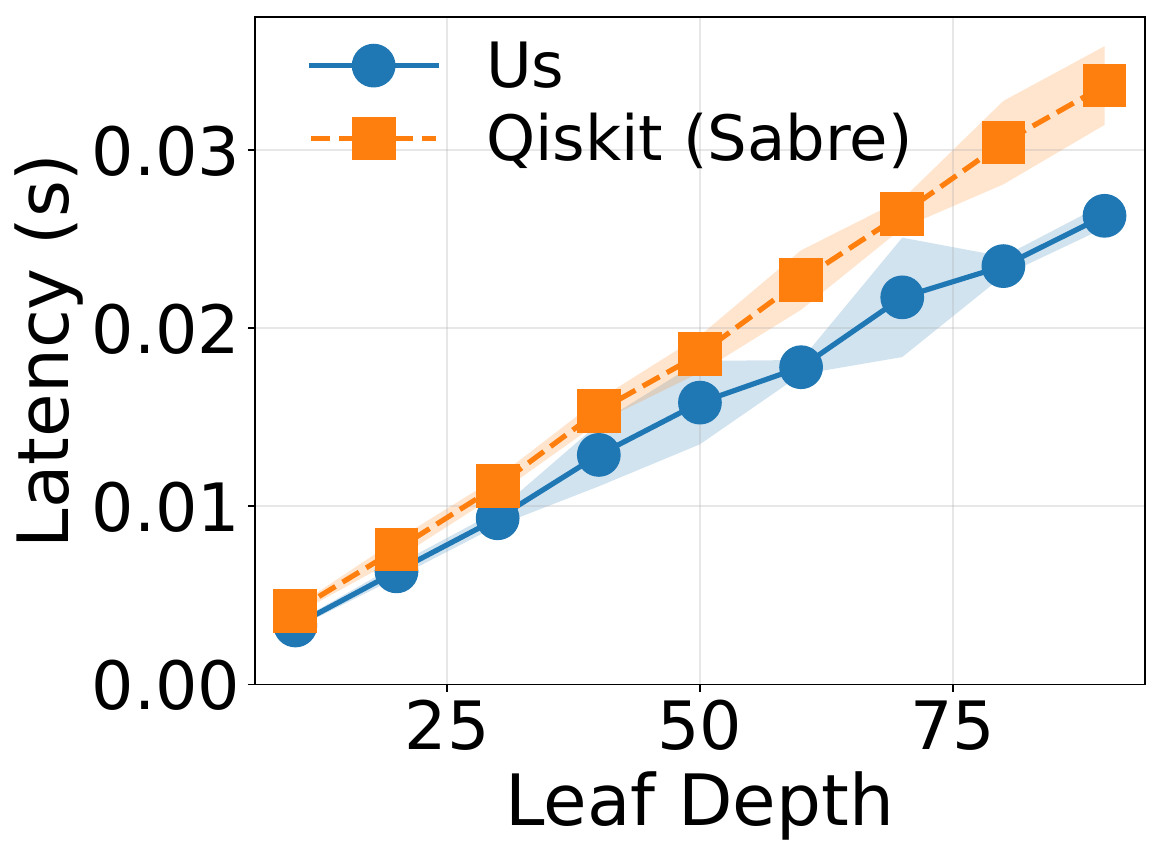}} &
        \subcaptionbox{\footnotesize 121qbt -- Error}[0.24\textwidth]{%
            \includegraphics[width=\linewidth]{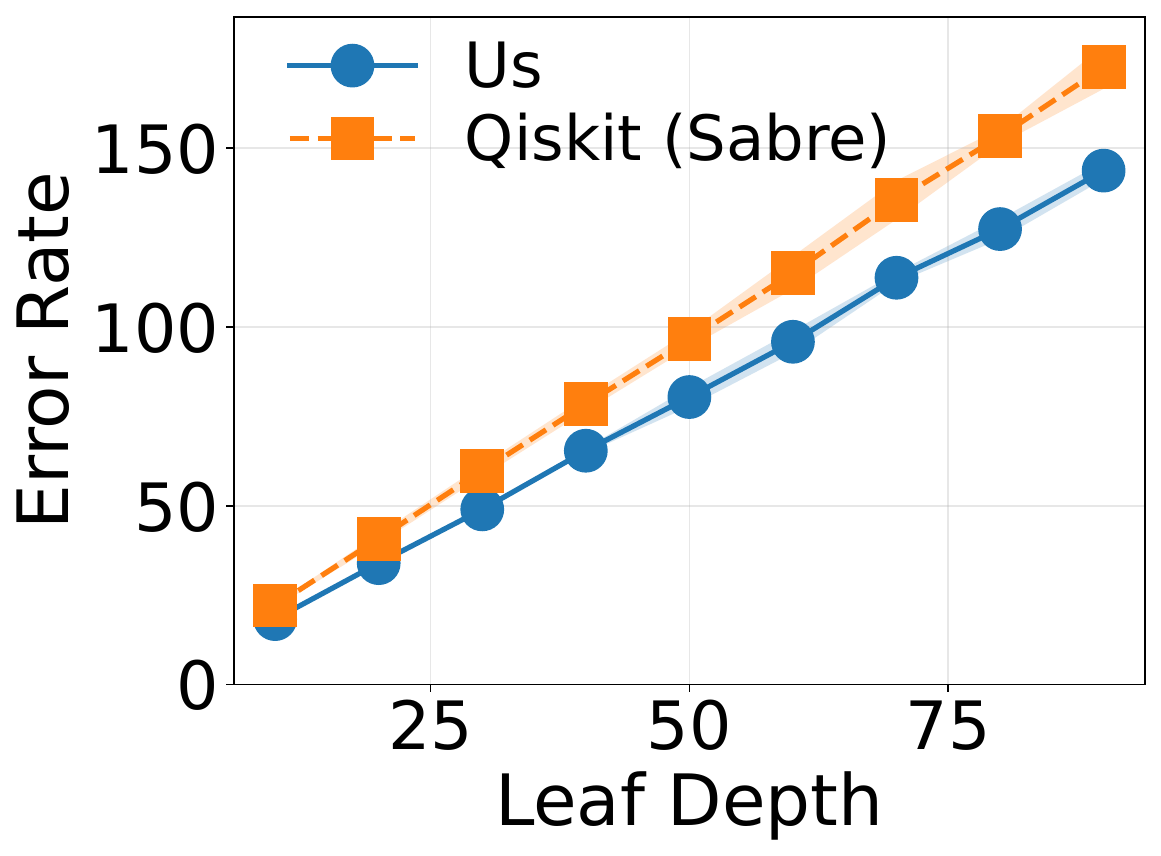}}
    \end{tabular}
    }
         \vspace{-2ex}
\caption{Routing performance on \textbf{IBM Brisbane} over 
\textbf{10 random
seeds} using the \textbf{Dynamic QUEKO benchmarks}. 
Plots show the mean
with $\pm1\sigma$ shaded (our method vs.\ Qiskit Sabre) across leaf depth. 
        Figures (a)--(d)  and (e)--(h) correspond
        to circuit widths, across the four metrics.
        X-axis (Leaf Depth) is the circuit depth of the static circuit in the innermost nested loop.
\vspace{-2ex}
        }
    \label{fig:results_dynamic_kingston}
    \Description{}
\end{figure*}

The summary of our results 
is shown
in Tab.~\ref{tab:results_dynamic_avg_improvements} while
Fig.~\ref{fig:results_dynamic_kingston} 
shows the absolute values for IBM Brisbane.
We compare with QISKIT LightSABRE, given
that it is the only one that supports dynamic circuits.
We first turn our attention to the high-level summaries
in 
Tab.~\ref{tab:results_dynamic_avg_improvements}.
We observe a marked improvement across both back-ends,
Brisbane (127 qubits) and Kingston (156 qubits).
The most notable improvements are achieved on the {\tt SWAP}
and {\tt Error} metrics.

\noindent
Improvements
are owed to the balance between
{\em weighted distance} ($C_\mathrm{spatial}$)
and
{\em depth-rate} ($C_\mathrm{rate}$)
terms 
(Sec.~\ref{subsec:cost_function}). 
The former term avoids
swap insertion across the critical path while
also avoiding noisy qubits and high-error links. Further, effects are compounded
due to the {\em reconciliation pass}
(Sec. \ref{subsec:reconciliation}) and {\em entry re-mapping}. 
In regard to the circuit {\tt Depth} and {\tt Latency}
metrics, \ourtool{} improves between 8.7\% and
18.59\% over Sabre. Overall, our results translate to faster
circuits that make use of 34\%--52\% fewer SWAP gates,
and between 14.62\%--40.53\% lower error.

\subsection{Evaluation on Chiplet-Based QPUs}\label{sec:chiplet-results-section}

We evaluate \textsc{DynamiQ} on chiplet-based QPUs using generated \textit{d-QUEKO} circuits, including a 256-qubit benchmark that stresses inter-chip communication ({\autoref{tab:chiplet-results}}).
On hexagon heavy $8{\times}8$\_$2{\times}2$ topologies (four chips in a
$2{\times}2$ grid, generated using MECH~\cite{zhang2024mech}), \textsc{DynamiQ}
consistently improves all metrics over Qiskit Sabre, reducing
\textsc{swap} count by 34--38\,\%, depth by 4--13\,\%, latency by 14--16\,\%,
and error by 10--12\,\%. 
On IBM Flamingo (three 156-qubit chips with a
$7.4{\times}$ inter-chip penalty), 
\textsc{DynamiQ} deliberately increases
\textsc{swap}s (+32\,\%) and depth (+20\,\%) to avoid costly inter-chip
communication. This trade-off reduces latency by 2.8\,\% and error by 15\,\%,
demonstrating effective hardware-aware optimization.


\begin{table}[H]
\centering
\caption{Routing results on chiplet-based QPUs: Heavy-Hexagon (Heavy H.) and IBM Flamingo (IBM F.)  using 81, 121 and 256 \textit{d-QUEKO} circuits.
Improvement denoted with $+P\%$, degradation denoted with $-P\%$.
\vspace{-2ex}
}
\label{tab:chiplet-results}
\resizebox{\linewidth}{!}{%
\begin{tabular}{ll rr rr rr rr}
\toprule
 &  & \multicolumn{2}{c}{\textbf{SWAPs ($\times 10^3$)}} 
   & \multicolumn{2}{c}{\textbf{Depth ($\times 10^3$)}} 
   & \multicolumn{2}{c}{\textbf{Latency ($\times 10^3$ $\mu$s)}} 
   & \multicolumn{2}{c}{\textbf{Error}} \\
\cmidrule(lr){3-4} \cmidrule(lr){5-6} \cmidrule(lr){7-8} \cmidrule(lr){9-10}
\textbf{Topology} & \textbf{Qubits}
  & Ours & Sabre
  & Ours & Sabre
  & Ours & Sabre
  & Ours & Sabre \\
\midrule
Heavy H. & 81  
  & \textbf{2.97} & 4.51 
  & \textbf{30.1} & 34.8 
  & \textbf{45.8} & 54.6 
  & \textbf{11.38} & 12.88 \\
Heavy H. & 121 
  & \textbf{4.57} & 7.41 
  & \textbf{33.0} & 34.4 
  & \textbf{50.4} & 58.3 
  & \textbf{12.31} & 13.65 \\
\cmidrule(lr){1-10}
\multicolumn{2}{c}{\textbf{Overall imp.}}
  & \multicolumn{2}{c}{\textbf{+36.2\%}}
  & \multicolumn{2}{c}{\textbf{+8.7\%}}
  & \multicolumn{2}{c}{\textbf{+14.9\%}}
  & \multicolumn{2}{c}{\textbf{+10.7\%}} \\
\specialrule{1.2pt}{2pt}{2pt}
IBM F. & 256 
  & 15.9 & \textbf{12.1} 
  & 40.9 & \textbf{34.2} 
  & \textbf{5122.0} & 5270.4 
  & \textbf{154.40} & 181.72 \\
\cmidrule(lr){1-10}
\multicolumn{2}{c}{\textbf{Overall imp.}}
  & \multicolumn{2}{c}{$-31.8\%$}
  & \multicolumn{2}{c}{$-19.6\%$}
  & \multicolumn{2}{c}{\textbf{+2.8\%}}
  & \multicolumn{2}{c}{\textbf{+15.0\%}} \\
\bottomrule
\end{tabular}%
}
\vspace{-2ex}
\end{table}






\subsection{Evaluating on Rotated Surface-Code Circuits}\label{sec:surface-code-results}

We evaluate \ourtool{} on rotated surface-code circuits generated using
\textsc{Stim}~\cite{gidney2021stim}.
Each circuit implements a memory-$Z$ experiment using $2d^{2}-1$ qubits
($d^{2}$ data + $d^{2}-1$ ancilla) with a four-step \textsc{cx} schedule. We consider two backends: IBM Brisbane (127 qubits, heavy-hex) and a
chiplet-based MECH~\cite{zhang2024mech} topology (480 qubits, $3{\times}4$ array).
For Brisbane, we sweep $d\in\{3,5,7\}$ and $r\in\{3,5,10,15,20\}$, where $d$ denotes the code distance and $r$ the number of syndrome extraction rounds, with
three replicates and three routing seeds (135 runs per method).
For MECH, we use $r{=}3$ with the same replicates and seeds (27 runs).

Table~\ref{tab:surface-merged} reports per-distance averages.
On IBM Brisbane, \textsc{DynamiQ} reduces SWAP count by 31.5\%, depth by 45.7\%, latency by 48.1\%, and error by 28.4\% on average.
On the MECH chiplet topology, it achieves similar gains, reducing SWAPs by 29.6\%, depth by 27.0\%, latency by 42.1\%, and error by 28.6\%. Overall, \textsc{DynamiQ} consistently improves all metrics across both backends. On chiplets, it further adapts routing to minimize costly cross-chip interactions,
leading to particularly strong latency and error reductions despite limited inter-chip connectivity.
\vspace{-1ex}

\begin{table}[H]
  \centering
  \caption{Performance metrics of DynamiQ vs.\ Qiskit Sabre on rotated surface-code circuits across backends.
  Summary row, {\tt Overall imp.}, shows relative (percentual)
  improvement with $+$P\% indicating improvements.
  \vspace{-2ex}
  }
  \label{tab:surface-merged}
  \setlength{\tabcolsep}{4pt}
  \resizebox{\linewidth}{!}{%
  \begin{tabular}{c c c c  rr  rr  rr  rr}
    \toprule
    & & & &
    \multicolumn{2}{c}{\textbf{SWAPs}} &
    \multicolumn{2}{c}{\textbf{Depth}} &
    \multicolumn{2}{c}{\textbf{Latency}} &
    \multicolumn{2}{c}{\textbf{Error}} \\
    \cmidrule(lr){5-6} \cmidrule(lr){7-8} \cmidrule(lr){9-10} \cmidrule(lr){11-12}
    Backend & $d$ & Qubits & CX/rnd &
    Ours & Sabre &
    Ours & Sabre &
    Ours & Sabre &
    Ours & Sabre \\
    \midrule

    \multirow{3}{*}{Brisbane}
    & 3 & 17 & 24  & \textbf{848}   & 1\,360  & \textbf{395} & 562   & \textbf{229} & 355   & \textbf{1.92} & 2.84 \\
    & 5 & 49 & 120 & \textbf{4\,320} & 4\,887 & \textbf{806} & 1\,132 & \textbf{494} & 711   & \textbf{3.80} & 4.47 \\
    & 7 & 97 & 168 & \textbf{8\,271} & 13\,387 & \textbf{909} & 2\,192 & \textbf{564} & 1\,415 & \textbf{5.05} & 7.71 \\

    \cmidrule(lr){1-12}
    \multicolumn{4}{c}{\textbf{Overall imp. $\Delta\%$}} &
    \multicolumn{2}{c}{\textbf{+31.5\%}} &
    \multicolumn{2}{c}{\textbf{+45.7\%}} &
    \multicolumn{2}{c}{\textbf{+48.1\%}} &
    \multicolumn{2}{c}{\textbf{+28.4\%}} \\

    \midrule

    \multirow{3}{*}{MECH}
    & 3 & 17 & 24  & \textbf{420}   & 480    & 200 & \textbf{185} & 124 & \textbf{112} & \textbf{0.09} & 0.10 \\
    & 5 & 49 & 120 & \textbf{1\,462} & 2\,380 & \textbf{316} & 446    & \textbf{472} & 737   & \textbf{0.17} & 0.27 \\
    & 7 & 97 & 168 & \textbf{4\,552} & 6\,284 & \textbf{460} & 706    & \textbf{526} & 1\,091 & \textbf{0.25} & 0.33 \\

    \cmidrule(lr){1-12}
    \multicolumn{4}{c}{\textbf{Overall imp. $\Delta\%$}} &
    \multicolumn{2}{c}{\textbf{+29.6\%}} &
    \multicolumn{2}{c}{\textbf{+27.0\%}} &
    \multicolumn{2}{c}{\textbf{+42.1\%}} &
    \multicolumn{2}{c}{\textbf{+28.6\%}} \\

    \bottomrule
  \end{tabular}%
  }
\end{table}

\begin{table}[htb]
\TableFontSize
\centering
\caption{Average mapping time (s) for \textbf{\ourtool{}} on \textbf{dynamic circuits} 
across three leaf-depth categories: Small $(10\text{--}30)$, Medium $(40\text{--}60)$, and Large $(70\text{--}90)$. Leaf Depth is the circuit depth of the static circuit in the innermost nested loop. 
}
\vspace{-2ex}
\label{tab:qroqi-horizontal-comparison}
\resizebox{.8\linewidth}{!}{
\begin{tabular}{c S S S S S S}
\toprule
\multirow{2}{*}{\textbf{Qubits}} &
\multicolumn{3}{c}{\textbf{IBM Kingston (156 Q)}} &
\multicolumn{3}{c}{\textbf{IBM Brisbane (127 Q)}} \\
\cmidrule(lr){2-4} \cmidrule(lr){5-7}
& \textbf{Small} & \textbf{Med.} & \textbf{Large} 
& \textbf{Small} & \textbf{Med.} & \textbf{Large} \\
\midrule


54  & 0.22 & 0.48 & 0.73 & 0.19 & 0.37 & 0.56 \\
81  & 0.38 & 0.69 & 1.01 & 0.34 & 0.64 & 0.97 \\
121 & 0.86 & 1.48 & 2.03 & 0.67 & 1.10 & 1.52 \\

\bottomrule
\end{tabular}}
\vspace{-2ex}
\end{table}

\subsection{Scalability of \ourtool{} Mapping}
\label{sec:eval:scalability}
Tab.~\ref{tab:qroqi-horizontal-comparison} 
presents the average mapping time of \ourtool{} on d-\textsc{QUEKO} dynamic circuits with 54, 81, and 121 qubits, across varying leaf-depth categories. \new{All mapping times were measured on a single core of an Intel Core i7-10750H (2.60\,GHz) with 8\,GB of RAM.}
In general, we observe 
a scalable mapping time where the number
of back-end qubits exerts a first order effect,
while the qubit footprint of the benchmark set
increases the time in
a linear fashion.
Additional results on the mapping time for the IBM Kingston and IBM Brisbane backends are in the supplemental material.

\begin{figure}[t]
\centering

\begin{subfigure}{0.49\linewidth}
    \centering
    \includegraphics[width=\linewidth]{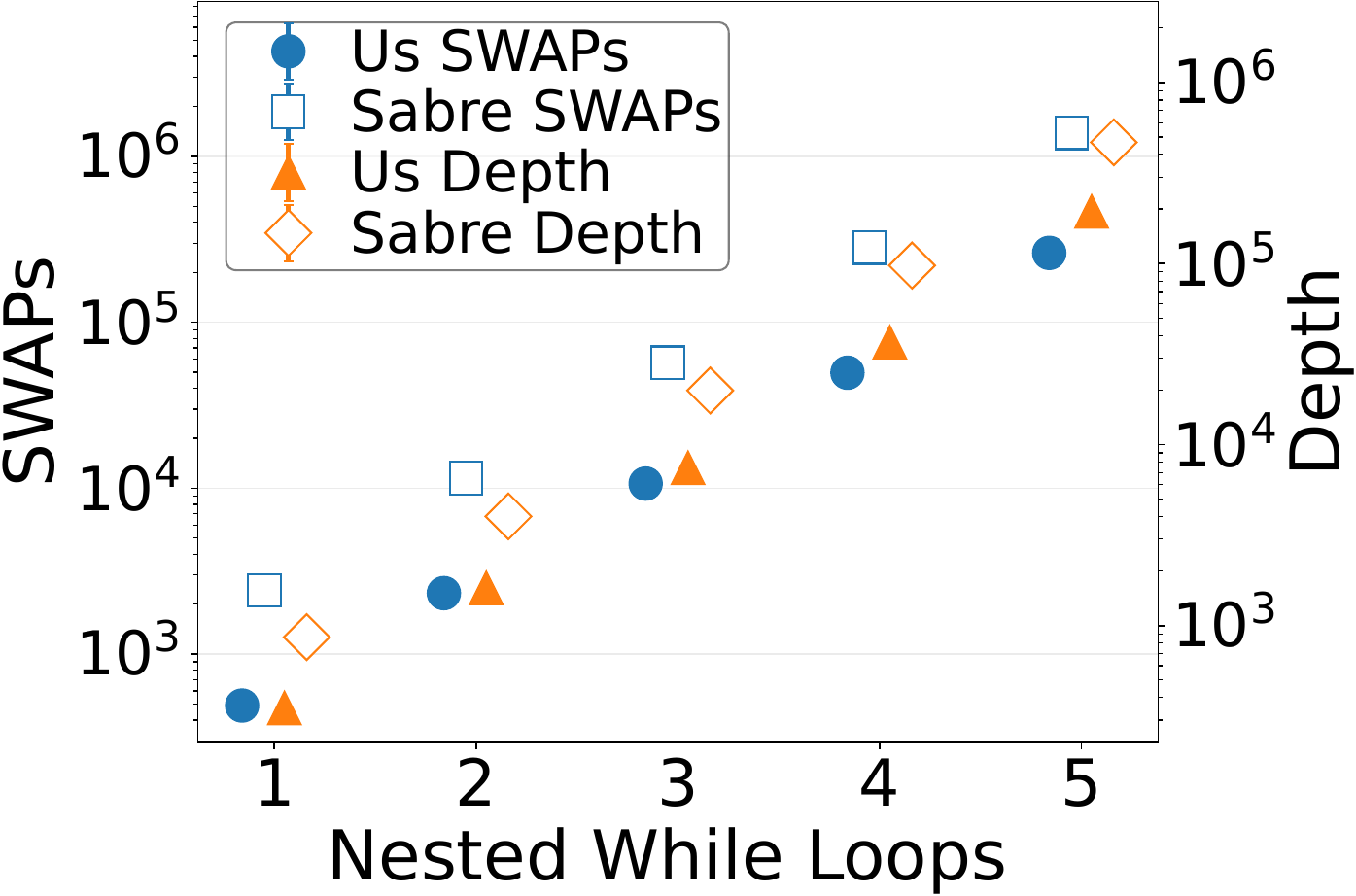}
    \vspace{-3.5ex}
    \begin{footnotesize}
    \caption{SWAPs and depth.}
    \end{footnotesize}
\end{subfigure}
\hfill
\begin{subfigure}{0.49\linewidth}
    \centering
    \includegraphics[width=\linewidth]{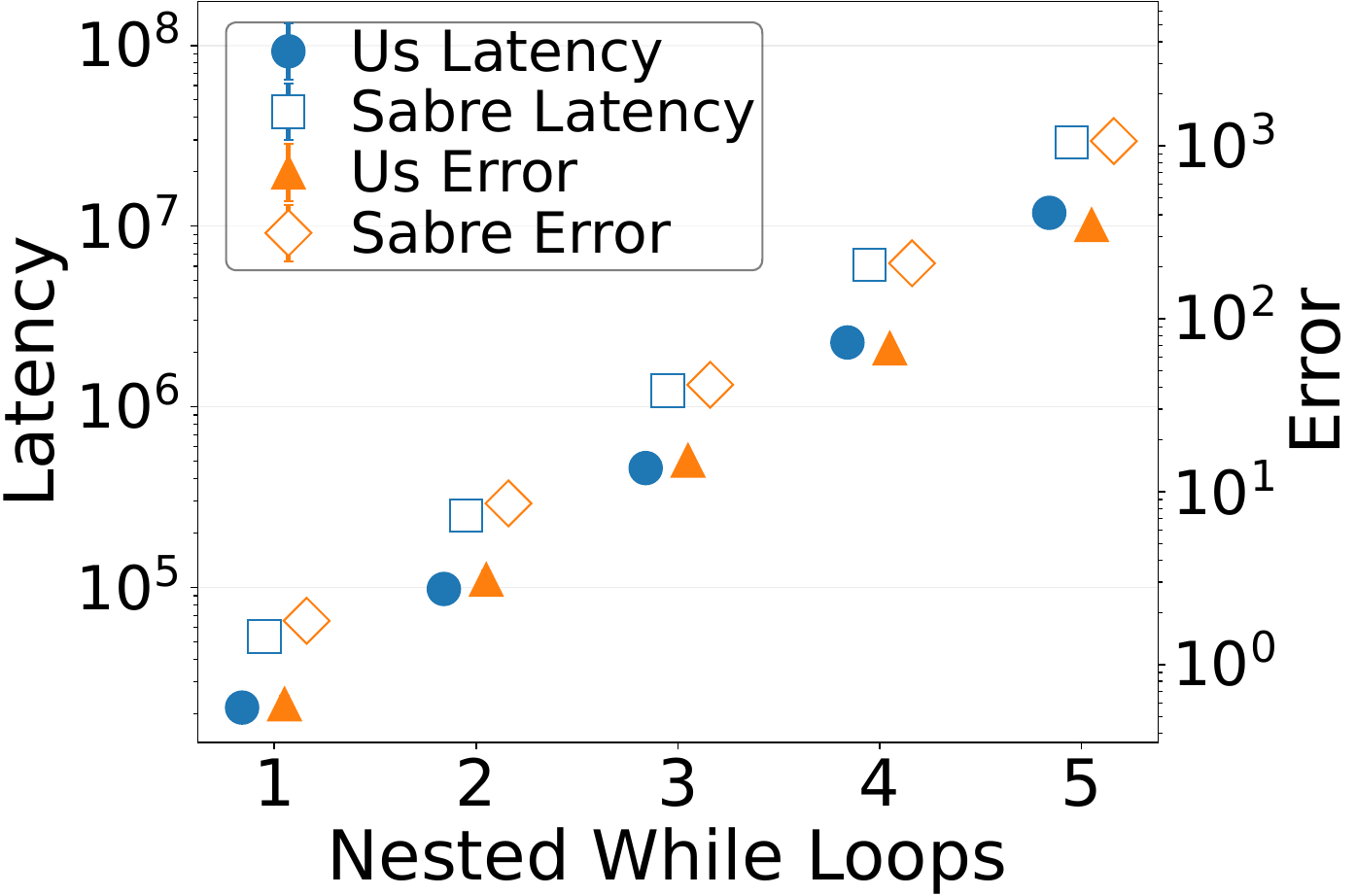}
    \vspace{-3.5ex}
    \begin{footnotesize}
    \caption{Latency and error rate.}
    \end{footnotesize}
\end{subfigure}

\vspace{-1.5ex}
\caption{
Dynamic nested-loop circuits on \(\mathbf{81}\)-qubit circuits, \texttt{ibm\_kingston}.
Four metrics shown on left and right Y-axes averaged over 3 random seeds and
reported as a function of the
number of nested \texttt{while} loops. 
Error bars omitted because of close proximity among values.
Lower is better.
}
\label{fig:nested_expirements_kingstone_wi_81q}
\Description{}
\end{figure}

\noindent
\begin{figure*}[tbhp]
    \shrink[1]
    \centering
    \setlength{\tabcolsep}{1pt} 
    \captionsetup[subfigure]{skip=0pt}
    \resizebox{0.98\textwidth}{!}{
    \begin{tabular}{@{}cccc@{}}

        \subcaptionbox{\footnotesize 81qbt -- SWAPs}[0.24\textwidth]{%
            \includegraphics[width=\linewidth]{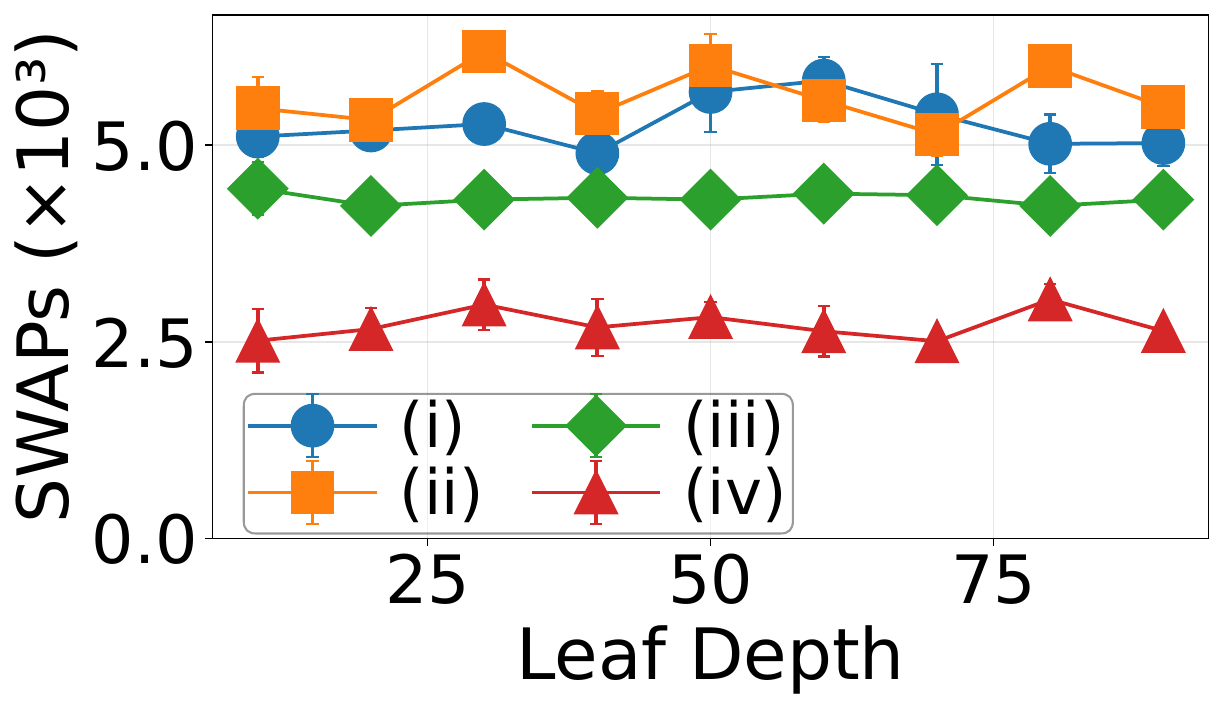}} &
        \subcaptionbox{\footnotesize 81qbt -- Depth}[0.24\textwidth]{%
            \includegraphics[width=\linewidth]{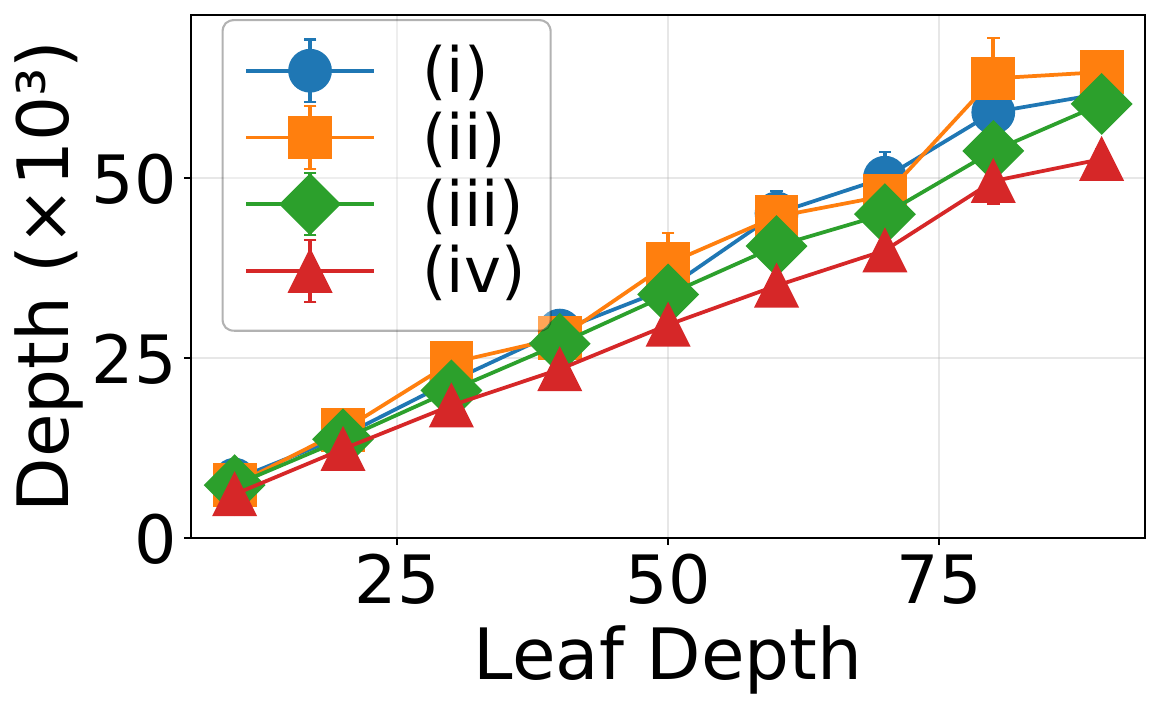}} &
        \subcaptionbox{\footnotesize 81qbt -- Latency}[0.24\textwidth]{%
            \includegraphics[width=\linewidth]{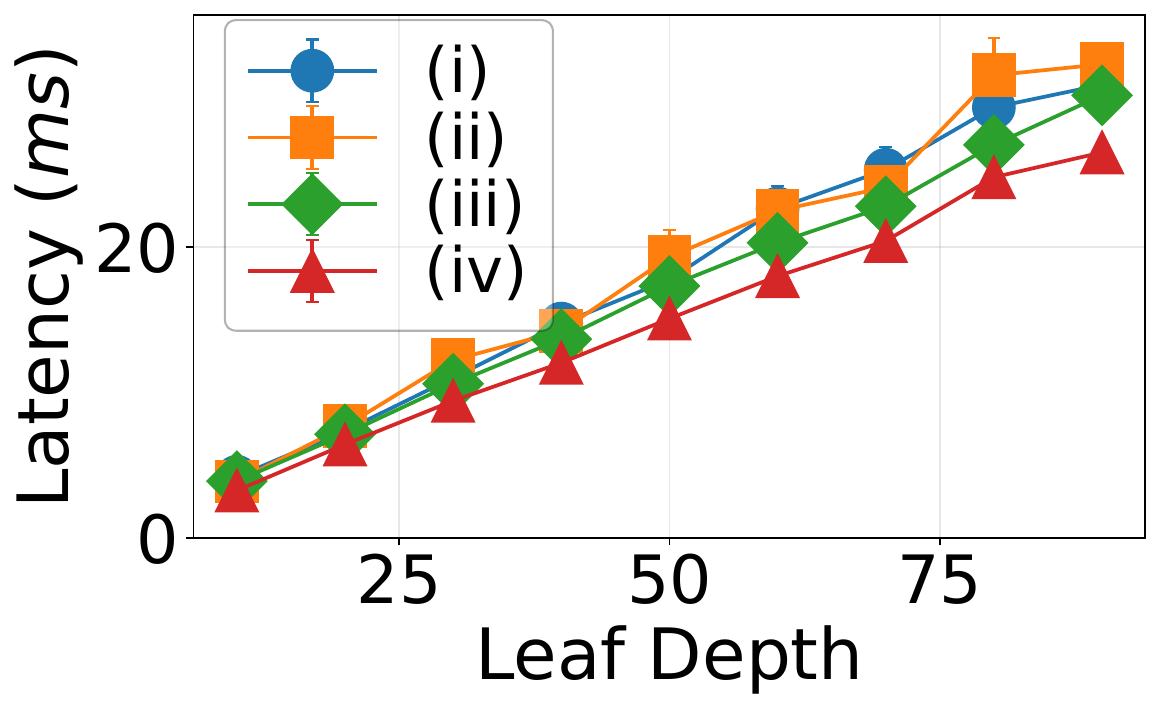}} &
        \subcaptionbox{\footnotesize 81qbt -- Error Rate}[0.24\textwidth]{%
            \includegraphics[width=\linewidth]{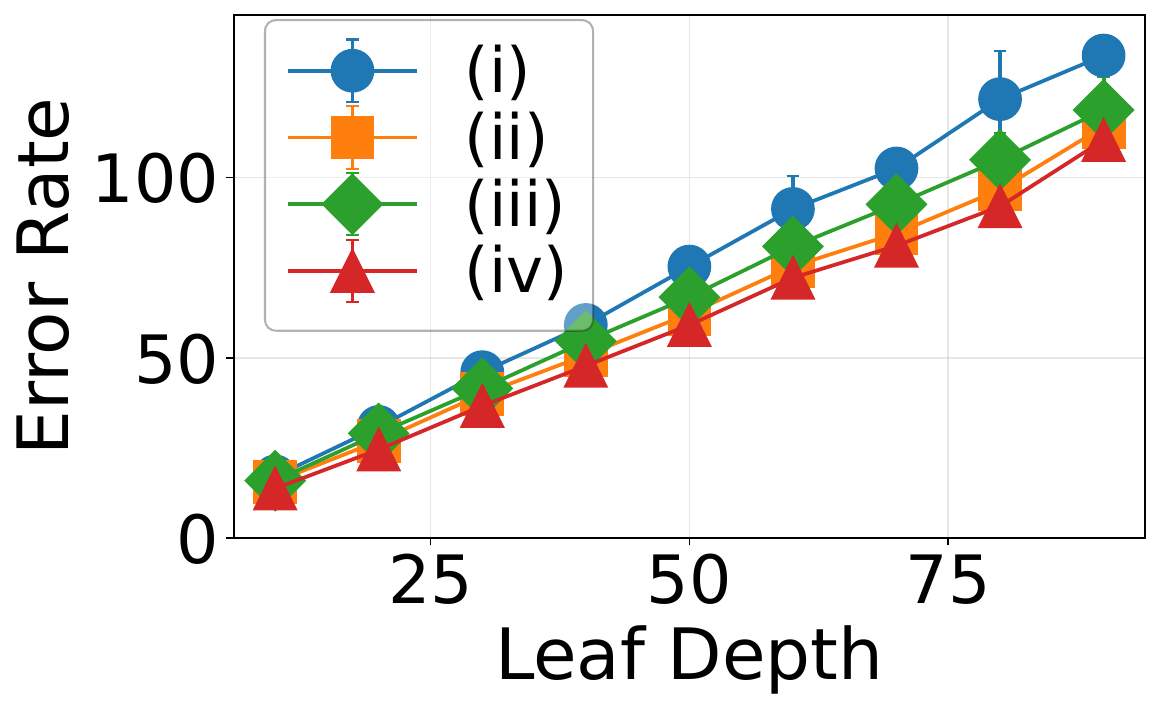}} \\[0.5em]

        \subcaptionbox{\footnotesize 121qbt -- SWAPs}[0.24\textwidth]{%
            \includegraphics[width=\linewidth]{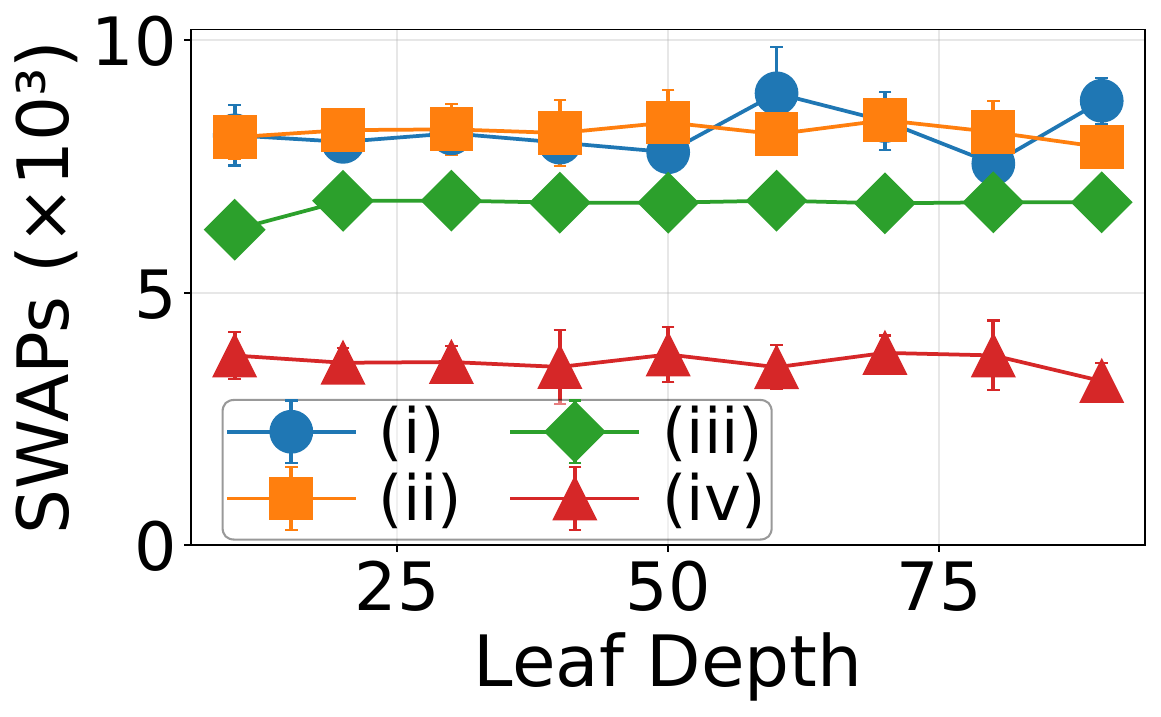}} &
        \subcaptionbox{\footnotesize 121qbt -- Depth}[0.24\textwidth]{%
            \includegraphics[width=\linewidth]{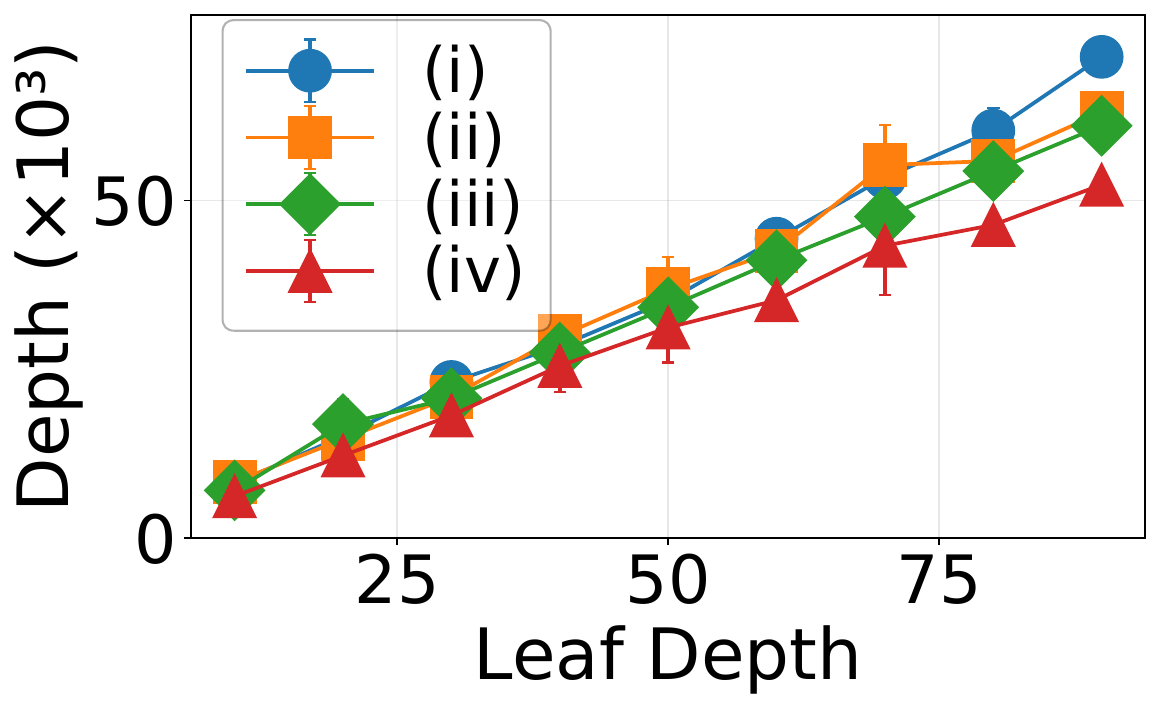}} &
        \subcaptionbox{\footnotesize 121qbt -- Latency}[0.24\textwidth]{%
            \includegraphics[width=\linewidth]{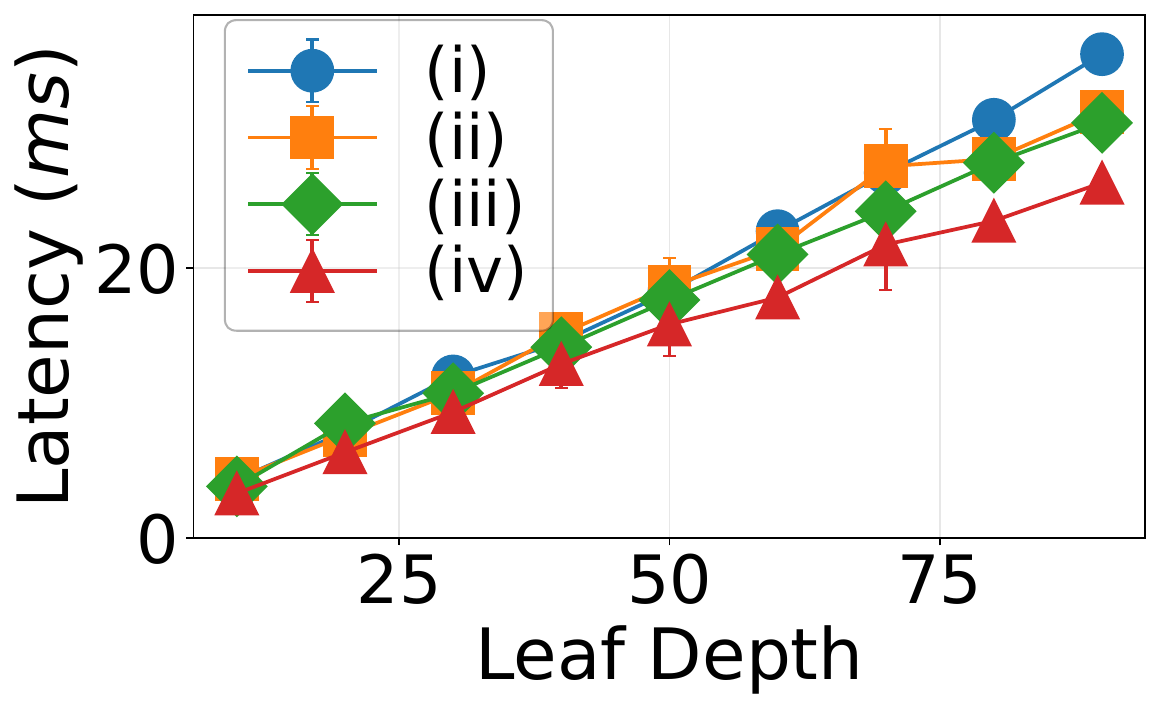}} &
        \subcaptionbox{\footnotesize 121qbt -- Error Rate}[0.24\textwidth]{%
            \includegraphics[width=\linewidth]{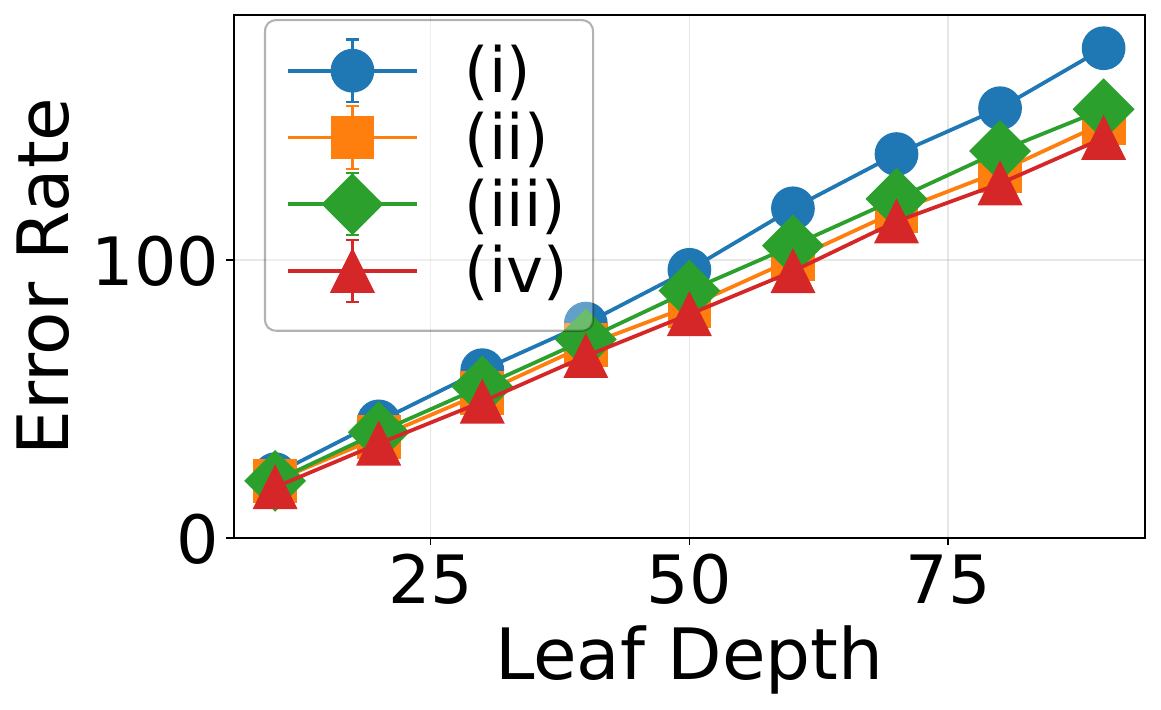}}
    \end{tabular}
    }
    \shrink[2]
    \caption{
        {\bf Ablation study} on  \textsc{Brisbane} back-end.
        Each panel shows mean values with $\pm1\sigma$ error bars across leaf depths over
        \textbf{10 random seeds}.
        Four levels are considered:
        (i) \textbf{Recon} (reconciliation only),
        (ii) \textbf{Recon+Err} (reconciliation with error awareness in the cost function),
        (iii) \textbf{Recon+Err+Depth} (error-aware reconciliation with additional depth regularization),
        and 
        (iv) \textbf{Full method} (the complete \ourtool{} model with all components enabled).
        Figures (a)--(d)  and (e)--(h) correspond
        to circuit widths, across the four metrics.
        X-axis (Leaf Depth) is the circuit depth of the static circuit in the innermost nested loop.
        \vspace{-3ex}
    }
    \label{fig:dynamic_brisbane_ablation}
    \Description{}
    \vspace{1ex}
\end{figure*}

\begin{table}[htbp]
    \TableFontSize
\vspace{-2ex}
\begin{footnotesize}
    \caption{Average percentage improvement (in~\%)  over the reconciliation-only baseline on dynamic circuits. Configurations are cumulative: (i) Recon, (ii) Recon+Err, (iii) Recon+Err+Depth, and (iv) Recon+Err+Depth+Remap. Metrics are SWAP count (S), depth (D), latency (L), and error (E). Positive values indicate better performance.
    }
\label{tab:dynamic_ablation_improvement}
\end{footnotesize}

\vspace{-2ex}
\setlength{\tabcolsep}{3pt}
\renewcommand{\arraystretch}{0.9}

\resizebox{\linewidth}{!}{
\begin{tabular}{
    l
    | *{4}{r}
    | *{4}{r}
}
\toprule
 & \multicolumn{4}{c|}{\textbf{81 Qubits}} 
 & \multicolumn{4}{c}{\textbf{121 Qubits}} \\
\cmidrule(lr){2-5} \cmidrule(lr){6-9}
\textbf{Method} 
& S & D & L & E 
& S & D & L & E \\
\midrule
(ii) Recon+Err 
& $-6.9$ & $-2.3$ & $-2.2$ & $15.1$
& $-0.3$ & $2.1$ & $3.5$ & $13.5$ \\
(iii) Recon+Err+Depth 
& $17.7$ & $6.5$ & $6.1$ & $9.7$
& $17.5$ & $6.6$ & $6.9$ & $10.0$ \\
(iv) Recon+Err+Depth+Remap 
& \textbf{48.2} & \textbf{17.7} & \textbf{17.1} & \textbf{20.5}
& \textbf{55.5} & \textbf{19.4} & \textbf{20.1} & \textbf{18.0} \\
\bottomrule
\end{tabular}
}  
\end{table}

\subsection{Ablation Studies}
\label{sec:exp:ablations}

\subsubsection{Impact of Loop Depth in Mapping Efficiency.}
\label{sec:exp:dynamic:loop-depth}
We 
first
explore the performance of \ourtool{}
on circuits with diverse hierarchical structure. 
Fig.~\ref{fig:nested_expirements_kingstone_wi_81q}
shows five classes of circuits with 1-5 nested loops,
each with 1-5 if-conditionals nested in them.
(Examples of circuits included 
in the supplemental material).
Metrics are computed assuming that all loops 
execute the full trip-count (10 iterations).
For conditional structures, we 
adopt a worst-case analysis among all of the branches.
As 
shown in
Fig.~\ref{fig:nested_expirements_kingstone_wi_81q}, \ourtool{} achieves significant gains across evaluated metrics,
with the circuit complexity  primarily increasing 
with the depth of while loops.
On average, we observe an \textbf{80\% reduction in SWAP count} and \textbf{50–60\% improvement} in latency, error, and circuit depth compared to Qiskit’s SABRE routing algorithm.

\subsubsection{What Drives Improvement on Dynamic Circuits}
\label{sec:exp:ablation-dynamic}


To 
attribute the contributions
of individual 
mapping components,
we perform an ablation study. 
Components are
added one by one.
Tab.~\ref{tab:dynamic_ablation_improvement} reports the average percentage
improvement of each configuration over the baseline, and
Fig.~\ref{fig:dynamic_brisbane_ablation} shows the corresponding trends across
metrics. The baseline, \emph{Recon} (i), uses only the reconciliation passes
to preserve mapping consistency across dynamic regions. 
It does not use
the error-aware term in the cost function, does not include the depth-rate term
$C_{\mathrm{rate}}$, and does not optimize loop-entry mappings through
steady-state selection. We then add components cumulatively: \emph{Recon+Err}
(ii) adds hardware-error awareness to the cost function; \emph{Recon+Err+Depth}
(iii) further adds the depth-rate term $C_{\mathrm{rate}}$; and
\emph{Recon+Err+Depth+Remap} (iv) additionally enables steady-state loop-entry
remapping through \textsc{SelectSteadyState}. Thus, configuration (iv)
corresponds to the full \ourtool{} algorithm.

Tab.~\ref{tab:dynamic_ablation_improvement}
reports improvements for four metrics: SWAP count (S), circuit depth (D),
latency (L), and error (E). Positive values indicate improvement over the
reconciliation-only baseline. The results show that adding error awareness
primarily improves E, adding $C_{\mathrm{rate}}$ improves D and L, and adding
steady-state remapping yields the largest overall gains by reducing repeated
routing overhead inside loops.

\begin{table}[H]
{%
\centering
\caption{Sensitivity to $\langle \alpha,\beta\rangle$ on heavy-hexagon 8x8-2x2 QPU, \texttt{queko-bss-54qbt}. Baseline is
\texttt{spatial\_only} $\langle\alpha,\beta\rangle=\langle1,0\rangle$. Improvement is computed as
$(\text{Baseline}-\text{Variant})/\text{Baseline}\times 100$ (positive is better).
Pair \textbf{$\langle 1.0, 0.2 \rangle$} (bottom row) is used in the evaluation.
\vspace{-2ex}
}
\label{tab:alpha-beta-sweep}
\TableFontSize
\begin{tabular}{cc r rrrr}
\toprule
$\alpha$ & $\beta$  & SWAPs (\%) & Depth (\%) & Latency (\%) & Error (\%) \\
\midrule
0.50 & 0.50 & -42.00 &  8.00 & -23.83 & -23.25 \\
0.60 & 0.40 & -21.98 &  9.82 &  -9.73 & -10.63 \\
0.80 & 0.20 & -11.47 & 11.35 &  -1.82 &  -5.35 \\
1.00 & 0.40 & -11.19 & {\bf 16.99} &   2.11 &  -4.06 \\
{\bf 1.00} & {\bf 0.20} &  {\bf -2.61} & 12.40 &   {\bf 6.04} &  {\bf 0.35} \\
\bottomrule
\end{tabular}
}
\end{table}

\subsubsection{Ablation Study of the Cost Model Parameters}
\label{sec:alpha-beta-sensitivity}
We evaluate the sensitivity of \textsc{DynamiQ}'s cost model to the 
$(\alpha,\beta)$ parameters (Eq.~\ref{eq:M_s}), which balance
communication ($C_{\text{spatial}}$) and 
depth ($C_{\text{rate}}$)
objectives. 
Tab.~ \ref{tab:alpha-beta-sweep}
shows a shallow sweep of $\langle \alpha,\beta\rangle$ pairs;
it highlights the impact relative to the spatial-only
mode,
$\langle \alpha,\beta \rangle = \langle 1,0 \rangle $.
We observe that pairs with $\alpha < 1$ consistently
degrade performance
while
larger $\alpha$ prioritize reducing communication-driven costs (distance/error).
Increasing $\beta$ improves depth, but at the expense of higher
SWAP count and latency, reflecting a trade-off between short-term scheduling and
long-term communication overhead. 
(NOTE: All previous experiments use $\langle \alpha,\beta \rangle =\langle 1.0,0.2 \rangle$)

\vspace{-2ex}
{\subsection{\old{Evaluation on Static Circuits}}
\label{eval:static:main-paper}
\old{
Due to space limitations and to the main subject
of this work, we defer results on static circuits
to supplemental material.
}


\section{Related Work}
\label{sec:related}
{\bf Dynamic Circuits.}
Niu et al. introduce a suite of synthesis techniques 
for dynamic quantum
circuits, integrated into the BQSKit framework
\cite{Niu.dac.2024,niu2024acdcautomatedcompilationdynamic}. 
Their approach leverages unitary decomposition
techniques and topology-specific numerical synthesis 
to prepare arbitrary states prior to the 
qubit mapping passes.
On the other hand, QISKIT relies on specialized
control-flow constructs that simplify capturing quantum gates as sub-circuits, which are later
transpiled.
\new{
Chen et al. propose {\tt Sword}~\cite{new01-dynamic-circuits.tccad.2026}, which
segments circuits 
into a main and controlled sub-circuits,
each mapped using qubit occupancy and regional gate density.
Wall-clock time is used as objective, and each sub-circuit is assumed to execute once.
Sang et al.~\cite{new02-learning-mappings.aps.2026} allocate qubits to cores in
modular architectures with a learned policy, using mid-circuit measurement for
qubit reuse rather than control-flow, so the circuit structure remains static.
Cubeddu et al.~\cite{new03-mapping-unstructured-control-flow.arxiv.2019} 
address
unstructured control-flow, weighting basic blocks by expected execution count and
restoring placement consistency with inverse SWAPs placed according to the
dominator relation~\cite{general-dominators.popl.1992}. 
Their goal of a unique
placement matches ours, although we obtain it by selecting a
steady-state entry mapping per region rather than undoing each SWAP.
Ryan et al.~\cite{new04-hw-dynamic-circuits.2017} 
describe the control
electronics behind dynamic execution, confirming that the constructs we
target are realizable within qubit coherence times.
{\tt TeleSABRE}~\cite{new05-telesabre-teleport.tqe.2025}
extends SABRE with qubit and gate teleportation for inter-core movement
alongside intra-core SWAPs. They
harness new primitives
to support routing static DAGs, 
whereas \ourtool{} consumes
the program structure.
{\tt QGo}~\cite{new06-synthesis.icrc.2021} 
splits 
mapped circuits into small blocks
to re-synthesize them and reduce CNOT count.
Lourens et al.~
\cite{new07-hierarchical-circuit-neural-search.nature.2023}
build Quantum CNN architectures by nesting reusable circuit blocks (convolution and
pooling layers) across several levels, evidence that quantum circuits are naturally
written hierarchically.
Cheng et al.~
\cite{new08-2D-nearest-neighbor.2020}
perform 
qubit mapping
on 2D-grids 
by 
reducing a
weighted Manhattan distance 
frequency of qubit pair interactions. 
It then inserts SWAPs using a priority queue.
Annechini et al.~
\cite{new12-depth-driven-routing.dac.2025} scores SWAPs 
by their scheduling
depth rather than by path length, cutting routed depth by up to 70\%.  but targets routing of purely static circuits.
}

{\bf Affine-Based Mappers.}
Qlosure \cite{qlosure} introduces a dependence driven method
for scalable qubit mapping in static circuits based
on the transitive closure of dependence relations.
Crucially, it 
only supported
static circuits and driven solely by the circuit depth objective and inter-qubit distance.
In contrast, our work targets dynamic, hierarchical circuits
across 
error, latency, SWAP count, and circuit depth.

{\bf Qubit Reuse} \cite{Hua.asplos.2023,Huang.pldi.2024,Ding.isca.2020}.
This method involves resetting a 
qubit once its quantum state 
is
collapsed by measurement. 
After this,
the qubit can be
recycled 
to
avoid 
using new 
ancilla qubits, thereby reducing the circuit width. 
Nonetheless, qubit reuse is 
orthogonal to the problem of qubit mapping in both static and dynamic circuits.
QRCC \cite{Pawar.asplos.2024} analyzes qubit lifetimes using a qubit-reuse-aware DAG and formulates qubit reuse as an integer linear program (ILP), while Jiang et al.~\cite{Hanru.pldi.2024} use a qubit dependency graph (QDG) to analyze and validate recycling strategies. QR-Map~\cite{Kim.isca.2025} uses map-based abstractions to identify reuse opportunities. Qulin~\cite{Huang.pldi.2024} transforms 
conditional multi-qubit 
gates to a linear number of two-qubit operations without helper qubits, reducing code size while preserving correctness.

{\bf Uncomputation.}
Unqomp~\cite{Paradis.pldi.2021} 
automates 
uncomputation in circuits 
by returning helper qubits to the $|0\rangle$ state after use. SQUARE~\cite{Ding.isca.2020} 
optimizes the allocation and reclamation of ancilla qubits through uncomputation, balancing trade-offs among qubit, 
gate count, and parallelism. Venev et al.~\cite{venev-uncomputation.oopsla.2024} leverage 
a custom, modular IR to synthesize
uncomputation and adjoints.

\vspace{-3ex}
\section{Conclusion}
\label{sec:conclusion}
This paper introduces \ourtool{},
a new qubit mapping framework
for dynamic circuits.
\ourtool{} effectively
maps hierarchical quantum circuits 
exhibiting
data-dependent control-flow, which gives
rise to multiple sub-circuits.
We make four technical contributions.
First, we leverage the key concept of
{\em qubit reconciliation}, a pass that unifies
the qubit mapping state from different
control-flow paths, thereby enabling consistent
mappings across sub-circuit and hierarchical boundaries.
Second, we introduce  a novel block-entry remapping
that optimizes the qubit steady-state to which 
all control-flow paths must merge.
Third, we enhance a multi-objective cost function
based on transitive dependences, qubit depth-rate and
qubit error that allows us to optimize all four
circuit metrics.
Finally, we present the design and implementation of
{\tt d-QUEKO} (Dynamic QUEKO), a generator of dynamic
circuits exhibiting varying control structure.
We validate our approach on a wide
range of dynamic circuits on two monolithic
QPUs
of 127 and 156 qubits, and on chiplet hexagon-based QPUs.
On the former, DynamiQ 
improves the SWAP count by up to 52\%,
depth by up to 18\%, latency by up to 18.6\%,
and error by up to 40\%. On chiplet architectures,
we achieve improvements of up to 36\% on SWAP count,
8.7\% on depth, 15\% on latency, and 15\% of error.


\bibliographystyle{ACM-Reference-Format}
\bibliography{bib/martin,bib/quantum,bib/related,bib/dynamic,bib/applications,bib/token_swapping,bib/polyhedral}


\newcommand{\rebquest}[1]{\textcolor{magenta}{Q: #1}\\}
\newcommand{\rebansw}[1]{\textcolor{blue}{A: #1}\\}
\newcommand{\rebhint}[1]{\textcolor{violet}{H: #1}\\}

\end{document}